\RequirePackage{lineno} 
\documentclass[aps,prc,twocolumn,showpacs,superscriptaddress,nofootinbib]{revtex4-2} 
\DeclareMathAlphabet{\mathpzc}{OT1}{pzc}{m}{it}
\usepackage{lineno}

\usepackage[T1]{fontenc}

\usepackage{graphicx}
\usepackage{amssymb}
\usepackage{amsmath}
\usepackage{amsthm}
\usepackage[percent]{overpic}
\usepackage{float}
\usepackage{array}
\usepackage{multirow}
\usepackage{epsfig}
\usepackage{color}
\usepackage{verbatim}

\usepackage[hidelinks,colorlinks=true,allcolors=blue,urlcolor=blue]{hyperref}

\usepackage{simplemargins}
\settopmargin{1.8cm}
\setbottommargin{2.3cm}

\newcommand{\car}{\text{nat}}
\newcommand{\D}{\mathrm{d}}
\newcommand{\etof}{{E_\text{tof}}}
\newcommand{\en}{{E_n}}
\newcommand{\enq}{{E_n^2}}

\newcommand{\x}{\mathrm{x}}
\newcommand{\X}{\mathcal{X}}

\newcommand{\C}{\mathcal{C}}
\newcommand{\Pij}{\mathcal{P}}

\newcommand{\N}{\mathcal{N}}

\newcommand{\T}{\top}

\newcommand{\xyz}{{\color{white}^|}}

\newcommand{\vrt}{ {\color{white}\large |} }
\newcommand{\hrz}{}

\newcommand{\reff}{Ref.}
\newcommand{\reffs}{Refs.}

\begin{document}

\title{Energy-differential measurement of the $^\car$C(\textit{n,p}) and $^\car$C(\textit{n,d}) reactions at the n\_TOF facility at CERN}


\author{P.~\v{Z}ugec} \affiliation{Department of Physics, Faculty of Science, University of Zagreb, Zagreb, Croatia} %
\author{N.~Colonna} \affiliation{Istituto Nazionale di Fisica Nucleare, Sezione di Bari, Italy} %
\author{D.~Rochman} \affiliation{Paul Scherrer Institut (PSI), Villigen, Switzerland} %
\author{M.~Barbagallo} \affiliation{European Organization for Nuclear Research (CERN), Switzerland} \affiliation{Istituto Nazionale di Fisica Nucleare, Sezione di Bari, Italy} %
\author{J.~Andrzejewski} \affiliation{University of Lodz, Poland} %
\author{J.~Perkowski} \affiliation{University of Lodz, Poland} 
\author{A.~Ventura} \affiliation{Istituto Nazionale di Fisica Nucleare, Sezione di Bologna, Italy} %
\author{D.~Bosnar} \affiliation{Department of Physics, Faculty of Science, University of Zagreb, Zagreb, Croatia} %
\author{A.~Gawlik-Rami\k{e}ga } \affiliation{University of Lodz, Poland} %
\author{M.~Sabat\'{e}-Gilarte} \affiliation{Universidad de Sevilla, Spain} \affiliation{European Organization for Nuclear Research (CERN), Switzerland} %
% ...
%

\collaboration{The n\_TOF Collaboration (www.cern.ch/ntof)} \noaffiliation 

\begin{abstract}

Energy-differential cross section of the $^\car$C(\textit{n,p}) and $^\car$C(\textit{n,d}) reactions was measured at the neutron time of flight facility n\_TOF at CERN. The measurement was performed in the first experimental area (EAR1; flight path of 182.5~m). Two position-sensitive $\Delta E$-$E$ silicon telescopes were used. Two naturally occurring carbon isotopes, $^{12}$C and $^{13}$C, contribute to the reactions on natural carbon, with the (\textit{n,p}) reaction threshold at 13.7~MeV and the (\textit{n,d}) threshold at 14.9~MeV (determined by the $^{12}$C isotope for both reactions). This paper provides the details of the analysis leading to the final results published previously as a Letter. The cross section results are reported up to 25~MeV. During the data analysis the population of the excited states in the daughter nuclei $^{11}$B, $^{12}$B, $^{13}$B had to be considered, requiring the adoption of the branching ratios and angular distributions of the emitted particles from an external source of information. TALYS-2.0 calculations were used as the main source and an in-depth analysis of the model-related uncertainties was performed. The n\_TOF results are largely inconsistent with the major evaluation libraries. On the other hand, an unexpected agreement is found with TALYS-2.0 calculations. Specifically, the obtained cross section for the (\textit{n,p}) reaction is significantly higher than in the available evaluations, fully supporting the earlier finding from an integral measurement at n\_TOF.

\end{abstract}

\maketitle


\section{Introduction}

Neutron induced reactions on light nuclei play an important role in fundamental nuclear physics, medical applications and radiation detection techniques. The $^\car$C(\textit{n},cp) reactions with charged particles (cp) in the exit channel are especially relevant for their implications in hadrontherapy and detector development studies. As an abundant constituent of biological tissues, carbon is one of the most relevant elements to be considered in any form of hadrontherapy where the secondary neutrons are generated by the primary proton or light ion beams. The subsequent neutron induced reactions contribute to the unintended doses in healthy tissues, not only through the primary charged particles leaving a specific reaction, but also through a production and decay of secondary radioactive isotopes. For example, the (\textit{n,p}) reaction on both naturally occurring stable carbon isotopes -- $^{12}$C and $^{13}$C -- produces the short-lived $^{12}$B and $^{13}$B nuclei (half-lives of 20.2~ms and 17.3~ms) that undergo a $\beta$-decay, emitting the highly energetic electrons with the average energy of around 6~MeV. As such, the importance of the $^\car$C(\textit{n},cp) reactions in nuclear medicine has long since been recognized~\cite{chadwick_1996}, calling for a precise experimental investigation of their cross sections.

These reactions also affect the shielding and activation calculations in the radiation protection and space dosimetry, wherein the design and optimization of shielding materials exposed to intense neutron fields depends on the accurate knowledge of their cross sections. This is specially true for the  accelerator-based neutron facilities, spallation neutron sources and neutron irradiation facilities for fusion related material research, such as the IFMIF-DONES facility currently under construction~\cite{dones1,dones2}. In parallel, the development, use and calibration of the neutron sensitive diamond detectors all critically depend on a precise characterization of the neutron induced reactions responsible for the detector signals, especially in the fast-neutrons range around and above 10~MeV.

The broad relevance of these reactions has already motivated a number of experimental investigations of the (\textit{n,p}) or (\textit{n,d}) reaction; on $^{12}$C, $^{13}$C or $^\car$C; close to the reaction threshold or at higher energies~\cite{
kreger_1959,
rimmer_1968,
ablesimov_1971,
bobyr_1972,
pillon_2011,
pillon_2017,
kellogg_1953,
mcnaughton_1975,
jackson_1988,
brady_1991,
sorenson_1992,
yang_1993,
pourang_1993,
olsson_1993,
martoff_1996,
wang_1996,
slypen_2000,
majerle_2020,
kuvin_2021,
chinese_2021,
wantz_2025}. In some cases the partial cross sections were extracted, relating only to a few specific low-lying states in the daughter nuclei. Despite all these measurements, the cross sections within a few MeV above the reaction threshold remain poorly constrained, as the experimental data show considerable inconsistencies (see Fig.~\ref{fig_exfor} below). The experimental discrepancies are reflected in the cross section evaluations that in some cases differ significantly between nuclear data libraries (see Figs.~\ref{fig_13c} and~\ref{fig_tendl} below). The same can be said about the theoretical predictions based on different reaction models, some of which are implemented in the particle transport codes such as Geant4~\cite{geant4_1,geant4_2,geant4_3}. These models -- actively used by the community through such codes -- also yield large deviations in the predicted cross sections (see, for example, Fig.~4 in \reff~\cite{carbon_epja}), undermining the confidence in their predictive modelling.

These inconsistencies clearly illustrate both the challenges in theoretical modelling of the neutron induced reactions on carbon and the technical difficulties in measuring their cross sections. The integral cross section measurement from n\_TOF exacerbated the problem even further, by obtaining an integral cross section notably higher than predicted by any available evaluation~\cite{carbon_epja,carbon_prc}. This has motivated a more sophisticated, energy-differential measurement of the $^\car$C(\textit{n,p}) and $^\car$C(\textit{n,d}) reactions at n\_TOF, its main results having been reported as a Letter~\cite{carbon_letter}. In this paper we provide a detailed report on all aspects of the experimental setup, data acquisition, and the subsequent data analysis.

Section~\ref{experiment} provides the details on the n\_TOF facility and the used detector setup. Section~\ref{analysis} reports on the analysis of the raw experimental data, with a special emphasis on the procedure used for a discrimination of proton and deuteron counts from the relevant reactions. Section~\ref{reconstruction} lays out the mathematical modelling of these data, together with all procedures leading to a reconstruction of the underlying cross sections. Section~\ref{results} presents the final results and their interpretation. Section~\ref{conclusions} summarizes the main conclusions of this work. Appendix addresses the estimation of an overall detection efficiency.

\section{Experimental setup}
\label{experiment}

\subsection{n\_TOF facility}
\label{intro_ntof}

The neutron time of flight facility n\_TOF at CERN~\cite{ntof_0,ntof_recent} is one of the most luminous white neutron sources in the world. The neutron beam is produced by a 20~GeV proton beam from the CERN Proton Synchrotron. The experiment was performed during the third phase (Phase-3) of the operation of n\_TOF facility, during which an average of \mbox{$7\times10^{12}$} protons were delivered per pulse. At n\_TOF a 7~ns wide (RMS) pulsed proton beam with the minimum repetition period of 1.2 s irradiates a massive lead target. A spallation of lead nuclei is a source of initially fast neutrons (on average, about 300 neutrons per incident proton) and other neutral and charged particles.


The beamlines connect the spallation target to three experimental areas. The first experimental area (EAR1) is located at a horizontal distance of approximately 185~m from the target, the second experimental area (EAR2) at a vertical distance of 20~m from the target. The third experimental area (NEAR) has been constructed recently, at a short horizontal distance of 1.5~m from the target. During the transport toward EAR1 or EAR2, the majority of charged particles are swept from the beam by strong electromagnets placed along the beamline. The rest of ultrarelativistic spallation products reaching these two experimental areas form an intense burst known as the $\gamma$-flash, affecting the detectors and serving as a reference point for the time calibration of the incoming neutrons. Due to the proximity to a spallation target and a designated purpose of NEAR station for activation rather than for time of flight measurements, a beamline toward NEAR does not have a sweeping magnet.

Initially fast neutrons are moderated passing through a spallation target itself, as well as through layers of demineralized and borated water from a cooling system surrounding the target. The end product is a white neutron spectrum spanning the energy range from thermal ($\sim$10~meV) up to a few~GeV in case of EAR1. The neutron flux in EAR1 and EAR2 is experimentally well characterized \cite{flux_ear1,flux_ear2} and the beam production and transport mechanisms are well understood \cite{beam1,beam2}. A more detailed description of EAR1 and the general features of the n\_TOF facility may be found in \reff~\cite{ntof}; EAR2 is well documented in \reffs~\cite{ear2_1,ear2_2,ear2_3}; NEAR station is described in \reffs~\cite{near_1,near_2,near_3}.

\subsection{Detector setup}
\label{setup}

Energy differential measurement of the $^\car$C(\textit{n,p}) and $^\car$C(\textit{n,d}) reactions was performed at EAR1 of the n\_TOF facility. The choice of this experimental area is essentially related to the high neutron energies~$\en$ involved in the measurement ($\en$-thresholds above 10~MeV for all relevant reactions). Using a time of flight technique, a satisfactory neutron energy resolution at high energies may be achieved by using a long neutron flight path -- a condition favoring EAR1 over EAR2. Around $\en$ of 20 MeV a dominant contribution to this energy resolution comes from the time resolution of the silicon detectors, determined as the time spread between the signals in coincidence in the two silicon layers ($\Delta E$ and $E$). The RMS of their time differences of around 20~ns (see Fig.~7 from Ref.~\cite{sync}), in combination with the 7~ns spread of a proton beam from the Proton Synchrotron, leads to 1\% uncertainty in the reconstructed neutron energy.

A rectangular graphite ($^\car$C) sample of 5~cm~$\times$~5~cm in lateral dimensions and 0.25~mm thickness was placed in the neutron beam, covering the entire beam profile. Sample was tilted by 45$^\circ$ relative to the beam axis, thus the effective carbon thickness of 0.35~mm was exposed to the beam. This corresponds to an effective areal density of \mbox{$4\times10^{-3}$} atoms per barn. Carbon sample was surrounded by two identical silicon telescopes placed outside the beam -- one parallel to the beam and the other parallel to the sample -- covering the angular range from 20$^\circ$ to 140$^\circ$. Each telescope consists of two silicon layers, 5~cm~$\times$~5~cm wide and distanced by 7~mm. The first, $\Delta E$-layer, is 20~$\mu$m thin, while the second, $E$-layer, is 300~$\mu$m thick. Each layer is segmented into 16 strips (5~cm~$\times$~3~mm each), separated by a thin layer of inactive silicon. Strips are oriented in the same direction in both layers. Pictures and schematics of the experimental setup may be found in \reffs~\cite{carbon_letter,sync,angular,neural}.

Silicon telescopes allow for a charged particle discrimination by means of a $\Delta E$-$E$ technique. They have been well characterized~\cite{site_np} and previously used in a challenging measurement of the $^7$Be(\textit{n,p}) reaction~\cite{be_np}, relevant to as yet unresolved Cosmological Lithium Problem. Electronic signals from all 64 silicon strips were digitally recorded at 125~MS/s sampling rate, with 14-bit resolution. They were analyzed by the dedicated pulse shape fitting procedures developed and implemented at n\_TOF~\cite{psa}. The shape of the recorded signals is reported in \reffs~\cite{sync,site_np}.

\section{Data analysis}
\label{analysis}

Experimental data analysis consists of four basic steps: (1)~neutron energy calibration, (2)~deposited energy calibration, (3)~particle discrimination, (4)~cross section reconstruction. We describe each of these steps in the following sections. Two distinct types of energy will be of crucial importance to this work. These are the kinetic neutron energy and the energy deposited in the detectors by charged particles. Under the neutron energy we will further discriminate between the \textit{true} neutron energy~$\en$ and the \textit{reconstructed} neutron energy~$\etof$, calculated from the neutron time of flight. Under the deposited energy we will discriminate between the energies $\Delta E$ and $E$ deposited either in the thin or the thick layers from two silicon telescopes, i.e. in 20~$\mu$m thin $\Delta E$-strips or in 300~$\mu$m thick $E$-strips. 

Only a small portion of particles will reach the $E$-layer and be detected by the $\Delta E$-$E$ technique (well bellow 10\% for both protons and deuterons at neutron energies below 26~MeV). See Appendix for a discussion about the overall detection efficiency and the reason for selecting the 0.25~mm thick carbon sample.

\subsection{Neutron energy calibration}
\label{en_calibration}

The kinetic energy of neutrons inducing any of the reactions relevant to this work is determined by the time of flight technique. The technique consists in registering the detection time of the reaction products (all charged particles in this work) relative to the production time of the pulsed neutron beam, corresponding to the impinging of the proton beam on target. The time between the emission of reaction products and their detection is negligible relative to the neutron propagation time along the evacuated beamline leading from a spallation target. Hence, the time difference between the neutron production and the reaction product detection may be treated as a neutron time of flight~$T$ along the flight path of length~$L$. The (reconstructed) neutron energy~$\etof$ is then given via the relativistic relation:
\begin{linenomath}\begin{equation}
\textstyle \etof=\left\{\left[1-\left(\frac{L}{cT}\right)^2\right]^{-1/2}-1\right\}m_n c^2,
\label{etof}
\end{equation}\end{linenomath}
with the neutron mass $m_n$, and $c$ the speed of light in vacuum. The flight path of $L=182.5$~m between the spallation target and the experimental setup was obtained using the standard technique relying on known-energy capture resonances. In practice the reconstructed energy~$\etof$ does not necessarily correspond to a true neutron energy~$\en$, but is strongly correlated to it. This correlation is expressed through a so-called resolution function of the neutron beam~\cite{ntof_0,ntof,beam1,beam2,rf_ntof,rf_unfolding,rf_class} and will be discussed in detail in the following sections.

In determining the neutron time of flight a detector response to a $\gamma$-flash (see Section~\ref{intro_ntof}) is commonly used for an absolute time calibration. However, silicon telescopes used in this work feature a very weak response to a $\gamma$-flash (that being a design feature, rather than a flaw). For this reason an alternate source of synchronization was used and the Wall Current Monitor (WCM)~\cite{pkup} was selected for this purpose. WCM is an induction device registering the proton pulses delivered from the CERN Proton Synchrotron upon the spallation target, thus marking the starting point for the production of each neutron pulse. All 64 silicon strips were synchronized to this single time-reference source. To this end a special synchronization method was developed and used~\cite{sync}. Time offsets relative to a recorded WCM pulse were identified for each (\mbox{$i$-th}) strip separately, yielding similar values of approximately $\tau_i\approx300$~ns for all strips, related to the signal delay in propagating through the cables. The time of flight $T$ is then calculated as:
\begin{linenomath}\begin{equation}
T=t-t_\text{WCM}+L/c+\tau_i,
\end{equation}\end{linenomath}
with~$t$ the recorded time of the detected count and~$t_\text{WCM}$ the recorded time of the WCM pulse. The term $L/c$ accounts for a recorded WCM pulse being delayed due to a propagation of a $\gamma$-flash by a finite speed of light along a flight path of length~$L$.

\subsection{Deposited energy calibration}
\label{edep_calibration}


A 1.8~$\mu$m thin $^6$Li-enriched (95\%) LiF sample was used for the calibration of energy deposited by charged particles in the silicon strips. Placing a sample in the neutron beam, both the tritons and $\alpha$-particles from the $^6$Li(\textit{n},$\alpha$)\textit{t} reaction ($Q=4.78$~MeV) were detected. For the calibration purposes, only the data for \mbox{$\en<1$~keV} were used, so that the initial neutron energy does not affect the energy of emitted reaction products (2.73~MeV for tritons and 2.05~MeV for $\alpha$-particles).

The $\alpha$-particles were stopped in the 20~$\mu$m thin $\Delta E$-layer, thus providing a calibration reference only for $\Delta E$-strips. Most of the tritons punch through the $\Delta E$-layer and reach the $E$-layer, providing the spectra of counts detected in coincidence between the two layers (a time window of $\pm100$~ns was used for the identification of coincident pulses between the silicon layers~\cite{sync}).

Entire process was simulated by Geant4 particle transport code \cite{geant4_1,geant4_2,geant4_3} and the simulated results for all types of deposition spectra -- for tritons and $\alpha$-particles, in or out of coincidence -- were compared to the calibration measurements. Comparison between multiple types of energy deposition spectra allowed for a precise and reliable calibration of each strip's response to the charged particles.

\begin{table}[t!]
\caption{Coefficients parameterizing the neutron beam profile from Eq.~(\ref{profile}).}
\centering
\begin{tabular}{c|ccc|cc}
\hline\hline
&\hrz&  $\boldsymbol{\en<1\:\mathrm{keV}}$ $\xyz$ &\hrz&\hrz& $\boldsymbol{10\:\mathrm{MeV}<\en<30\:\mathrm{MeV}}$ $\xyz$\\
\hline
$\xyz \boldsymbol{A} \xyz$ && $38.0175\:\text{cm}^{-2}$ &&& $3.23117\:\text{cm}^{-2}$\\
$\xyz \boldsymbol{R} \xyz$ && $1.7\:\text{cm}$ &&& $1.55\:\text{cm}$\\
$\xyz \boldsymbol{k} \xyz$ && $0.08\:\text{cm}^{-2}$ &&& $0.65\:\text{cm}^{-2}$\\
$\xyz \boldsymbol{n} \xyz$ && $0.7$ &&& $3$\\
\hline\hline
\end{tabular}
\label{tab1}
\end{table}

\begin{figure}[t!]
\centering
\includegraphics[width=1\linewidth]{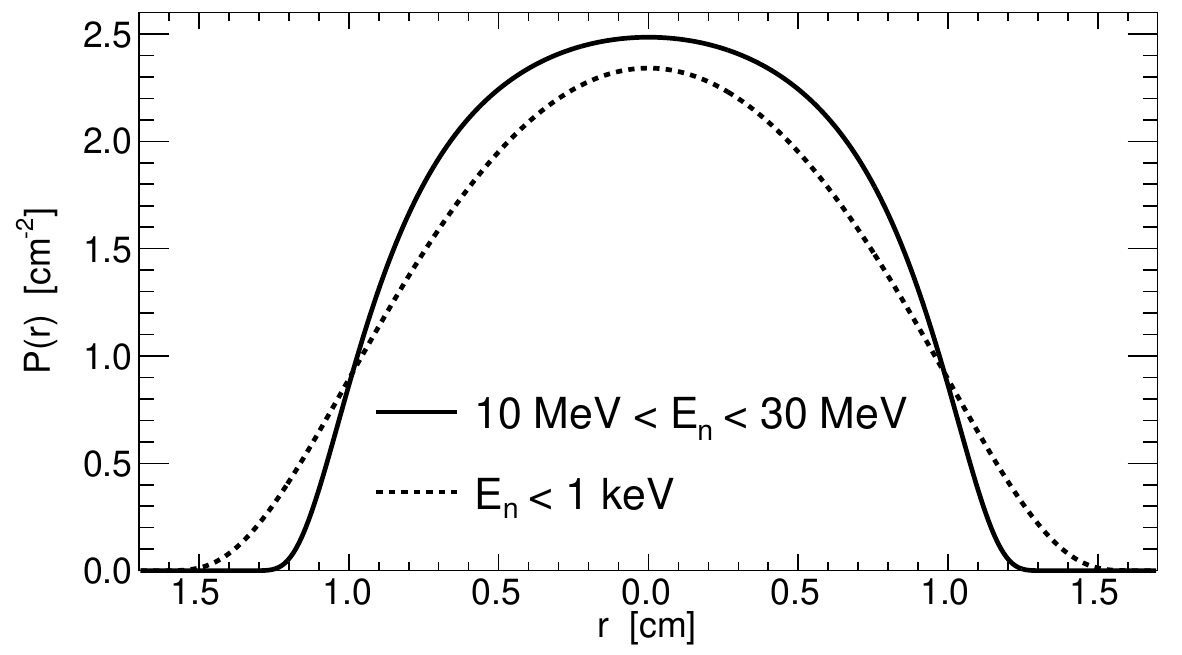}
\caption{Neutron beam profiles parameterized by coefficients from Table~\ref{tab1}.}
\label{fig_profile}
\end{figure}

In simulating the particle emission -- both for the calibration with LiF sample (for \mbox{$\en<1$~keV}) and for the $^\car$C data analysis at higher neutron energies (between approximately 10~MeV and 30~MeV) -- a special care was taken to accurately reproduce all aspects of the neutron beam, including its profile, i.e. a spatial distribution of neutrons around the beam axis. In general, this profile is neutron energy dependent and features a slight asymmetry around the beam axis. However, this asymmetry has no significant impact on our results. We will therefore treat the bean as axially symmetric, i.e. dependent only on the radial distance~$r$ from the beam axis. Let \mbox{$\D \Phi(\en)=\phi(\en)\D \en$} be a total number of neutrons of energy~$\en$ in the beam. Here $\phi(\en)$ represents the neutron flux. Furthermore, let $\D^2 \Phi(r,\en)$ be a total number of neutrons of energy $\en$ at the radial distance $r$ from the beam axis. The beam profile is then defined as:
\begin{linenomath}\begin{equation}
P(r; \en)\equiv \frac{\D^2 \Phi(r,\en)}{r\D r \times \D \Phi(\en)}=\frac{\D^2 \Phi(r,\en)}{r\D r \times \phi(\en)\D \en}.
\label{profile_def}
\end{equation}\end{linenomath}
At n\_TOF this quantity is determined from the dedicated FLUKA~\cite{fluka} and Geant4~\cite{geant4_1,geant4_2,geant4_3} simulations. Within a radius~$R$ the beam profile can be well described by the following distribution:
\begin{linenomath}\begin{equation}
P(r)=A\exp\left(-\frac{1}{[k(R^2-r^2)]^n}\right).
\label{profile}
\end{equation}\end{linenomath}
Outside this radius \mbox{$P(r)=0$}. In Eq.~(\ref{profile}) the coefficients $A$, $R$, $k$, $n$ are all neutron energy dependent, though we have dropped the $\en$ dependence for the clarity of notation. The coefficient $A$ is defined by a normalization condition \mbox{$\int_0^{R} P(r)r\D r=1$}. Within the energy range \mbox{$\en<1\;\text{keV}$} relevant for the deposited energy calibration with LiF sample, and within the range \mbox{$10\;\text{MeV}<\en<30\;\text{MeV}$} relevant for the $^\car$C data analysis the beam profile retains a stable form. Therefore, for each range a unique profile parametrization had to be used. The corresponding coefficients are reported in Table~\ref{tab1}. Figure~\ref{fig_profile} shows the difference in these two profiles.

\subsection{Particle discrimination}
\label{part_discrimination}

Once the neutron energy and the deposited energy calibration have been performed, each pair of pulses measured in coincidence between $\Delta E$ and $E$ layers is characterized by $\etof$ and a pair of deposited energies $\Delta E$ and $E$. We refer to each such pair of pulses as a \textit{coincident count}, that is uniquely represented by a point ($\etof,\Delta E,E$) in a three-dimensional parameter space. This allows for a particle discrimination, since different charged particles produce different patterns in this parameter space. While similar, these patterns are still unique to each $\Delta E$-$E$ pair of silicon strips. In practice they are rather nontrivial and those relevant to this work are very close in the parameter space, making an optimal separation between specific particle patterns a challenging issue. Machine learning techniques are well suited to this kind of classification problems. To this end a neural network based particle classification procedure was developed in \reff~\cite{neural}. A separate network was used for each relevant $\Delta E$-$E$ pair of silicon strips. Each neural network was trained on a carefully prepared and optimized dataset, obtained from the Geant4 simulations of the charged particle transport through the experimental setup. Protons, deuterons and tritons from $^\car$C(\textit{n,p}), $^\car$C(\textit{n,np}), $^\car$C(\textit{n,d}) and $^\car$C(\textit{n,t}) reactions were considered ($^\car$C consisting of $^{12}$C and $^{13}$C), since the protons and tritons from $^\car$C(\textit{n,np}) and $^\car$C(\textit{n,t}) reactions limit a discrimination of protons and deuterons from $^\car$C(\textit{n,p}) and $^\car$C(\textit{n,d}) reactions. Aside from the (\textit{n,np}) reactions on two carbon isotopes having a higher threshold than the (\textit{n,p}) reactions -- 17.3~MeV for $^{12}$C(\textit{n,np}) and 18.9~MeV for $^{13}$C(\textit{n,np}) in the laboratory frame -- the  (\textit{n,np}) reactions have three particles in the exit channel (\textit{n}, \textit{p} and a residual boron nucleus). For this reason the energy spectrum of the (\textit{n,np}) protons is continuous, similarly to the spectrum of electrons from a $\beta$-decay. This further reduces the average energy of the (\textit{n,np}) protons, when compared to those from the (\textit{n,p}) reactions. For this reason the (\textit{n,np}) protons do not interfere with a discrimination of the (\textit{n,p}) protons below 25~MeV~\cite{neural}.

\begin{figure*}[t!]
\centering
\includegraphics[width=0.39\linewidth]{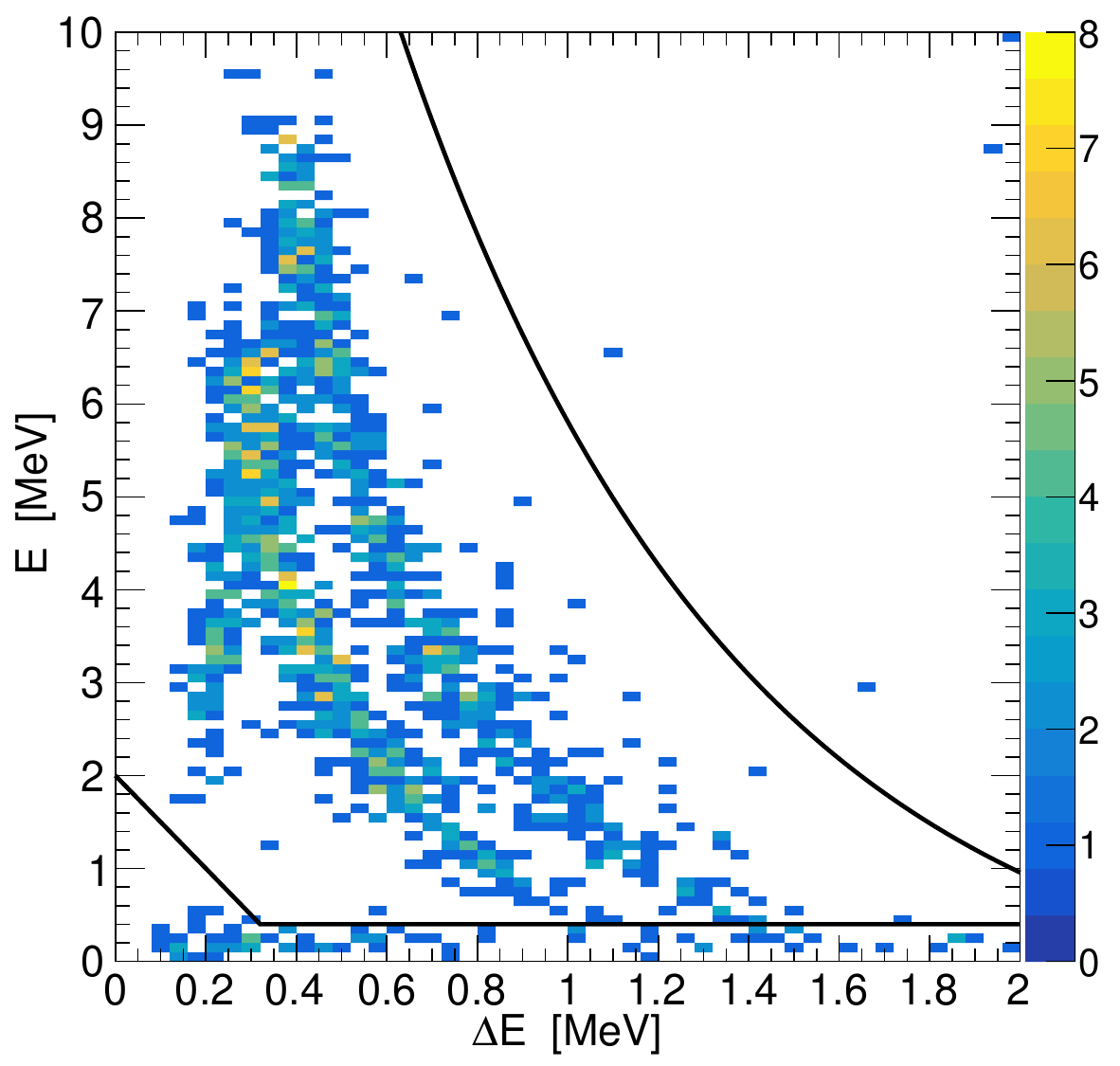}\includegraphics[width=0.39\linewidth]{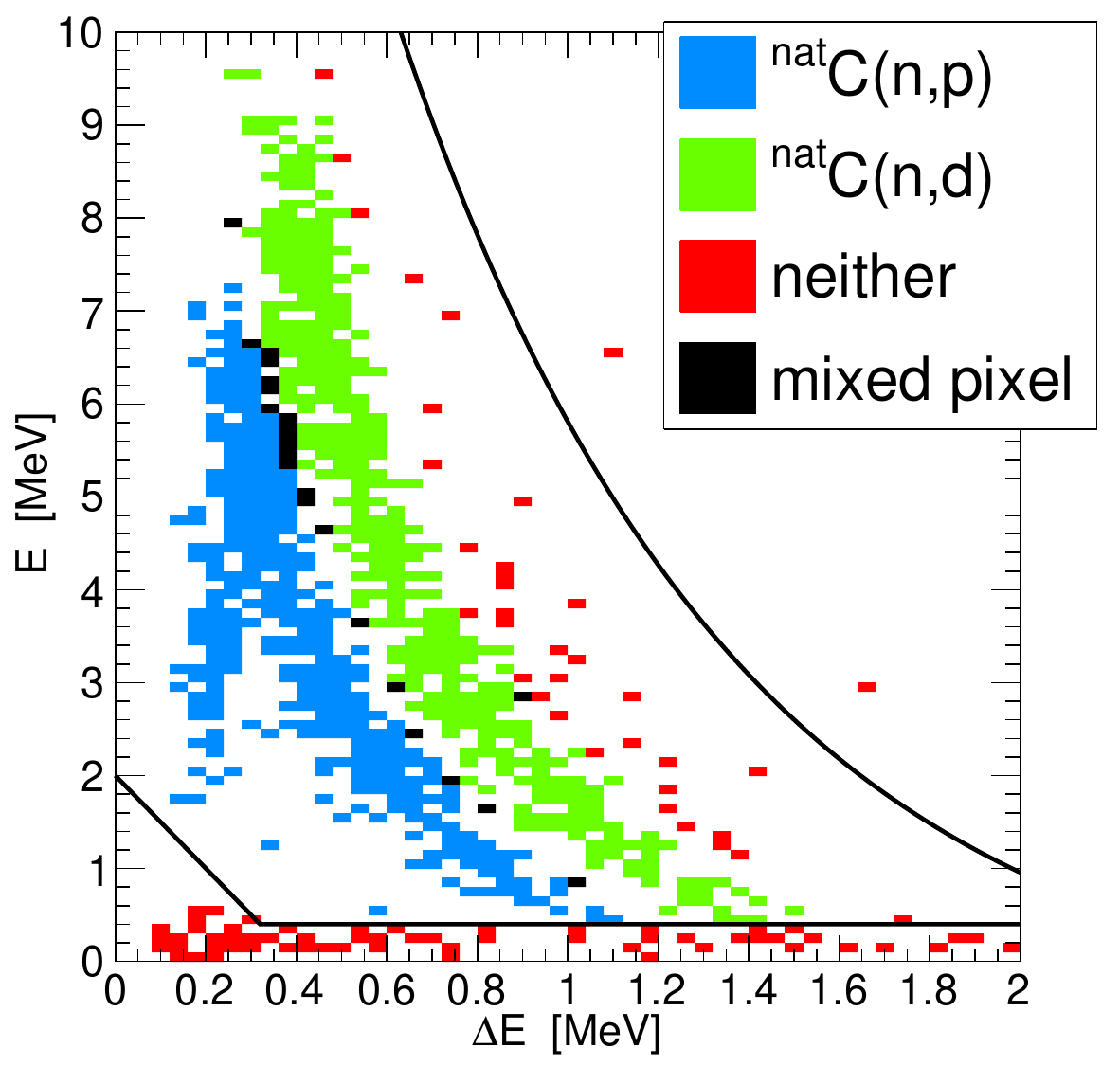}
\caption{Counts detected in coincidence by one arbitrary selected $\Delta E$-$E$ pair of silicon strips. All counts from the reconstructed neutron energy interval $14\;\text{MeV}<\etof<26\;\text{MeV}$ are shown. Right panel: raw counts; left panel: particle identification by trained neural network. See the main text for further details.} 
\label{fig_class}
\end{figure*}

In preparing the training datasets a care has been taken to apply both the deposited energy resolution and the reconstructed neutron energy resolution of the experimental setup to the simulated data. The deposited energy resolution has been experimentally determined: 2\% relative resolution (full width at half maximum) for $\Delta E$-strips and 0.5\% for $E$-strips~\cite{site_np}. The reconstructed neutron energy resolution, on the other hand, is a nontrivial property of the neutron beam. It is parameterized by the so-called resolution function~$R(\etof,\en)$, which intrinsically depends on the true neutron energy~$E_n$~\cite{ntof_0,ntof,beam1,beam2,rf_ntof,rf_unfolding,rf_class}. It is obtained from the dedicated simulations of the neutron production and their subsequent transport toward each experimental area, and is benchmarked against the available experimental data. While the resolution function is non-Gaussian for \mbox{$\en<1$~MeV}, between 4~MeV and 1~GeV it can be well modeled by an $\en$-dependent Gaussian distribution. Since the neutron energies relevant to this work are well within the 10~MeV -- 30~MeV interval, we have fitted the relevant portion of the resolution function to the following form:
\begin{linenomath}\begin{equation}
R(\etof,\en)=\frac{1}{\Delta(\en)\sqrt{2\pi}}\exp\left(-\frac{[\etof-\en-\delta(\en)]^2}{2\Delta^2(\en)}\right)
\label{rf_gaus}
\end{equation}\end{linenomath}
in order to facilitate its application. Within the 10~MeV -- 30~MeV interval the Gaussian shift~$\delta(\en)$ and width~$\Delta(\en)$ can be parameterized as:
\begin{linenomath}\begin{align}
&\delta(\en)=a_\delta \en+b_\delta,
\label{mali_delta}\\
&\Delta(\en)=a_\Delta \enq+b_\Delta \en+c_\Delta,
\label{veli_delta}
\end{align}\end{linenomath}
with: 
\begin{linenomath}\begin{gather*}
a_\delta=-0.00416,\, b_\delta=0.0075\,\text{MeV};\\
a_\Delta=9.55\times10^{-5}\,\text{MeV}^{-1},\, b_\Delta=0.0032,\, c_\Delta=-0.008\,\text{MeV}.
\end{gather*}\end{linenomath}
Given the number $\D \N(\en)$ of detected counts \textit{produced} by the neutrons of true energy~$E_n$, the resolution function translates this experimentally inaccessible quantity into experimentally accessible dependence $\D N(\etof)$ of counts \textit{detected} at the reconstructed energy $\etof$~\cite{rf_unfolding,rf_class}:
\begin{linenomath}\begin{equation}
\frac{\D N(\etof)}{\D\etof}=\int_0^\infty \D\en \left[R(\etof,\en)\frac{\D \N(\en)}{\D \en}\right].
\label{dn_detof}
\end{equation}\end{linenomath}
This is a basis for any computational procedure involving the resolution function.

By analyzing the separation between proton and deuteron patterns in~\cite{neural}, it was observed that the relevant reactions can not be discriminated according to the carbon isotope ($^{12}$C or $^{13}$C). Hence, we will treat the detected protons and deuterons as coming from the $^\car$C(\textit{n,p}) and $^\car$C(\textit{n,d}) reactions. It was also observed that these two particle types may be reliably discriminated from each other up to approximately 25~MeV in neutron energy. Above this energy both the protons and deuterons manage to punch through the thick $E$-layer of silicon strips, so that their $\Delta E$-$E$ patterns start to backbend and overlap. Due to the later considerations related to the energy binning and the resolution function of the neutron beam, in this work we consider the experimental data up to approximately 26~MeV in the reconstructed neutron energy. The lowest threshold for observation of coincident counts corresponding to the relevant reaction products is slightly above 14~MeV (see Section~\ref{linearizing}). Figure~\ref{fig_class} shows for an arbitrarily selected $\Delta E$-$E$ pair of silicon strips the pattern of detected counts within the reconstructed neutron energy range \mbox{$14\;\text{MeV}<\etof<26\;\text{MeV}$}. The left panel shows the raw counts prior to the particle classification. The right panel shows the particle classification by the trained neural network, wherein the proton and deuteron patterns are clearly visible. Black lines show the manual cuts fencing off the portions of parameter space that have been excluded from the data analysis, clearly representing unreliable or background events. Several counts beyond the curved black cut on the rightmost part of the plot correspond to $\alpha$-particles, which are well separated from the proton and deuteron patterns. The counts on the right side of the deuteron pattern, \textit{within} the black cuts, correspond to the tritons from the  $^\car$C(\textit{n,t}) reaction. Mixed pixels from the right panel are a byproduct of a finite pixel width used for visualization; each particular count is still uniquely assigned a specific particle type. Mixing of types within a given pixel may happen for two reasons: (1)~the type-separation boundary passes through a particular pixel and the experimental counts of both types are present within this pixel; (2)~the shape and the position of the type-separation boundary slightly varies with the neutron energy. Thus, counts with the same ($\Delta E,E$) coordinates may belong to different particles at different values of $\etof$.

\pagebreak

\begin{figure}[t!]
\centering
\begin{overpic}[width=1\linewidth]{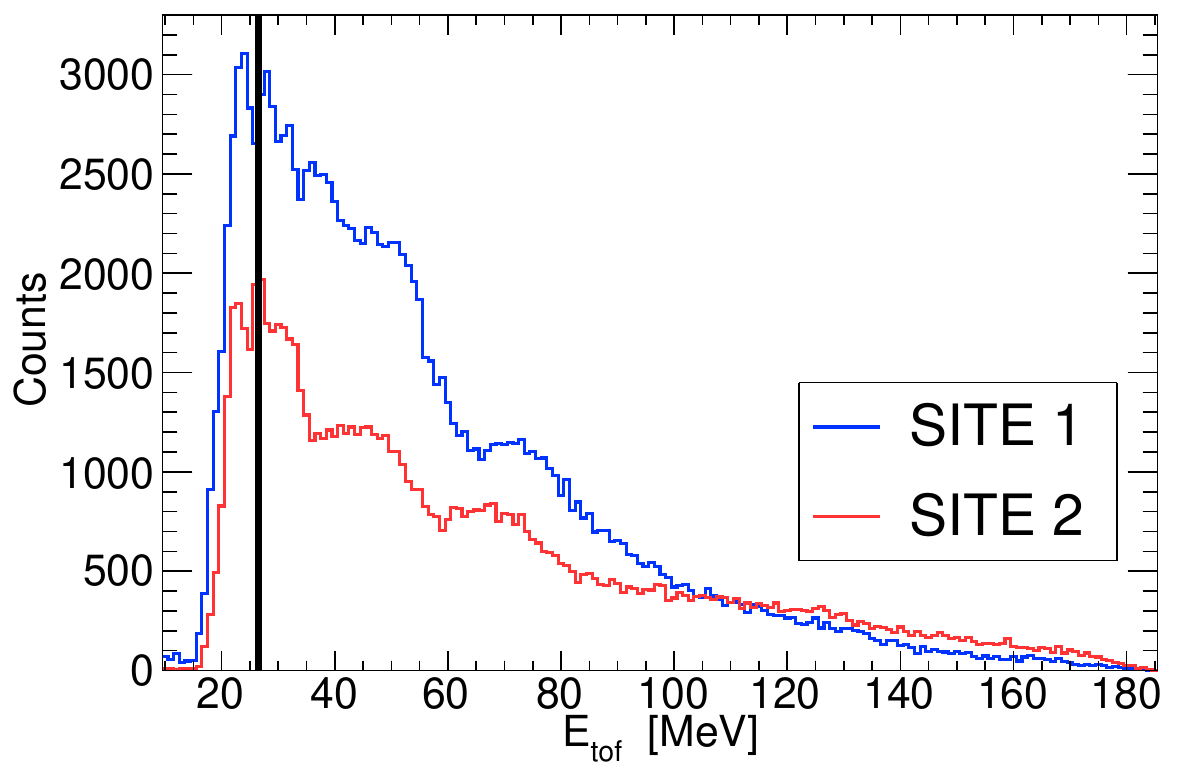}
\put(38.5,36){\includegraphics[width=0.57\linewidth]{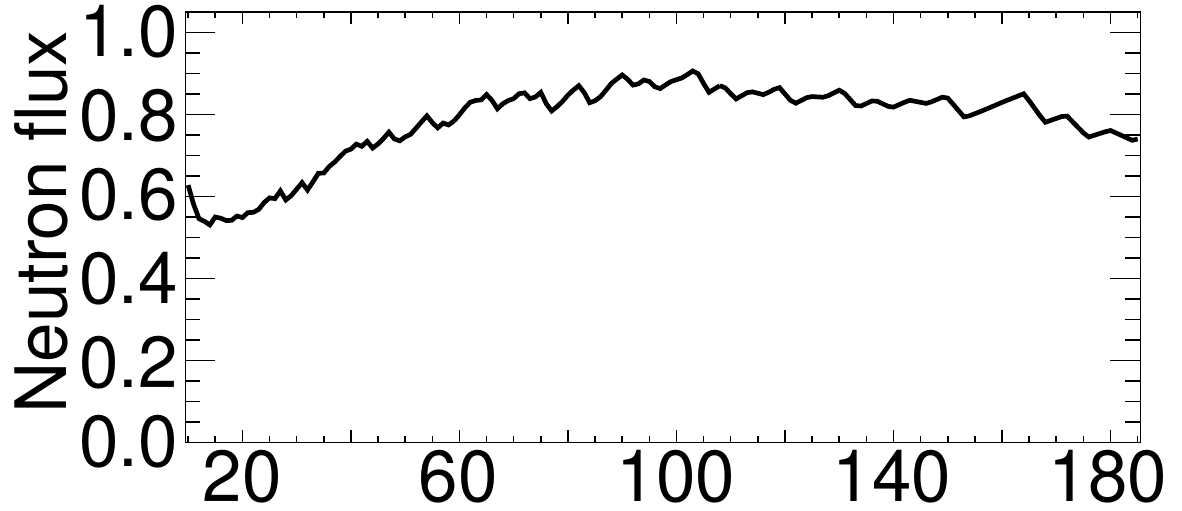}}
\end{overpic}
\vspace*{-5mm}
\caption{All counts (protons, deuterons and tritons) detected within the black lines from Fig.~\ref{fig_class}, registered by 34 relevant $\Delta E$-$E$ pairs of silicon strips from one silicon telescope (SITE~1; parallel to the sample, covering front angles) and 28 pairs from the other one (SITE~2; parallel to the beam, covering rear angles). Spectral structures are \textit{not} caused by the neutron flux, shown in the inset (flux in arbitrary units).}
\label{fig_all_counts}
\vspace*{-5mm}
\end{figure}

Figure~\ref{fig_all_counts} shows all coincident counts confined within the black lines from Fig.~\ref{fig_class}, detected by all 62 relevant $\Delta E$-$E$ pairs of silicon strips considered in the analysis. These 62 pairs -- which were selected for the sufficient statistics of detected counts -- comprise 34 pairs from one silicon telescope (SITE~1) and 28 pairs from the other one (SITE~2). The spectra show clear structures in their $\etof$ dependence and can be obtained up to a very high neutron energy from our measurement. These structures are \textit{not} caused by the shape of the neutron flux, which is shown by the inset in Fig.~\ref{fig_all_counts}. Therefore, these structures strongly indicate a complex behavior in the cross sections of the $^\car$C(\textit{n,p}), $^\car$C(\textit{n,d}) and/or $^\car$C(\textit{n,t}) reactions, which may warrant further experimental investigation.

\begin{figure}[t!]
\centering
\includegraphics[width=1\linewidth]{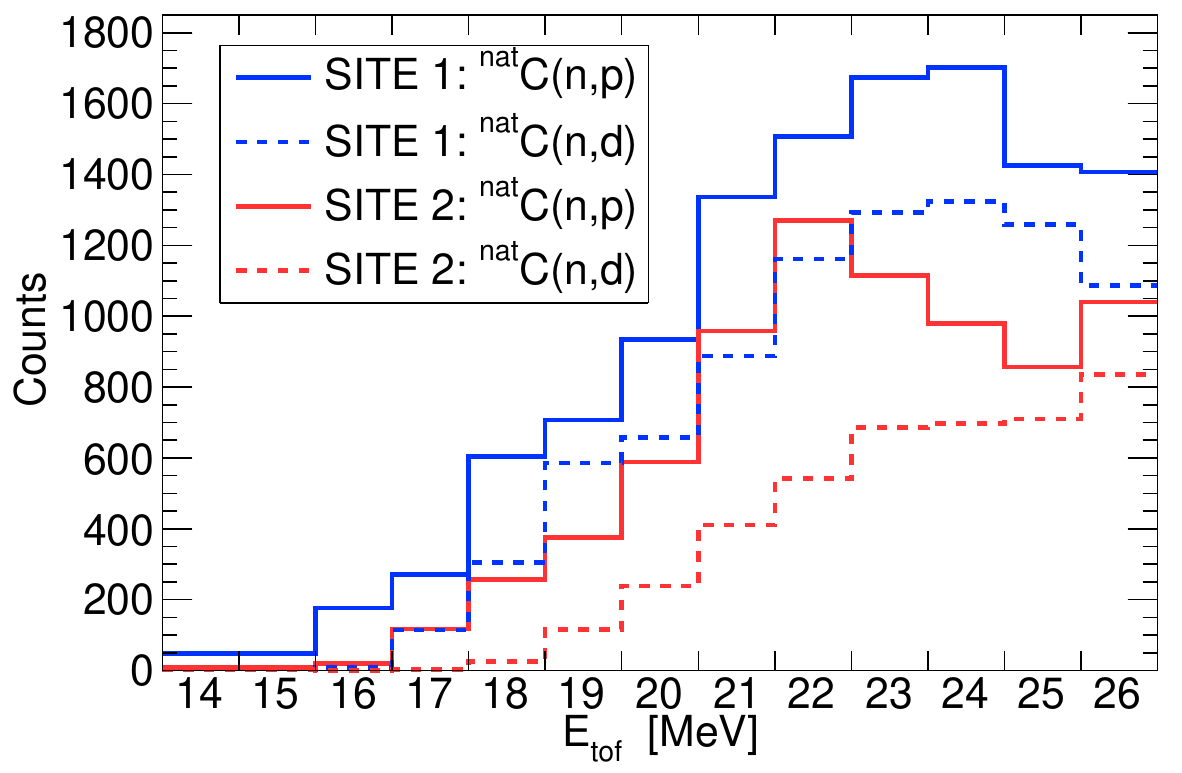}
\vspace*{-5mm}
\caption{Coincident proton and deuteron counts from two silicon telescopes, within the energy range where they can be reliably discriminated. See Fig.~\ref{fig_all_counts} for SITE distinction.}
\label{fig_nd_all}
\vspace*{-3mm}
\end{figure}

The vertical line in Fig.~\ref{fig_all_counts} separates the energy range relevant to this work ($\etof<26.5$~MeV; see below) from the range where particle types can no longer be reliably discriminated. Figure~\ref{fig_nd_all} focuses on the relevant range, showing separately the proton and deuteron counts detected by each silicon telescope. One reason for a difference in a detection yield between the two telescopes is the angular distribution of the reaction products. Another contribution comes from the differences in their neutron energy dependent detection efficiencies, caused by the reaction kinematics. At a given neutron energy, the backwards-emitted products (towards SITE~2) have a lower energy than the  forwards-emitted ones (towards SITE~1). In SITE~2 this increases the amount of low energy products stopped in the $\Delta E$-layer before reaching the $E$-layer, thus reducing a yield (see Appendix for a discussion on the energy dependent detection efficiency).

 Figures~\ref{fig_all_counts} and~\ref{fig_nd_all} are representative of the total statistics obtained during the measurement. They also roughly indicate the general shape of the cross sections. The actual data that will be used in the cross section reconstruction are not these total counts but rather the separate counts from each relevant $\Delta E$-$E$ pair of silicon strips, presented in Fig.~\ref{fig_counts} below.

\section{Cross section reconstruction}
\label{reconstruction}

\subsection{Modeling the observable data}
\label{modeling}

One of the major challenges in extracting the reaction cross sections from the measured data is related to the anisotropic angular distribution of the reaction products. Furthermore, for neutron energies up to 25~MeV multiple excited states may be populated in the residual boron nuclei, making the angular distributions dependent on these states. Therefore, the branching ratios for the accessible excited states also need to be taken into account. For the $^\car$C sample used one also needs to consider its isotopic composition consisting of 98.9\% of $^{12}$C and 1.1\% of $^{13}$C, since the detected proton and deuteron counts can not be discriminated according to a particular carbon isotope.

Evidently, a detailed mathematical model is needed in order to translate the cross sections into observable counts. This model was already developed in~\cite{angular}, not accounting for the resolution function of the neutron beam. As such, the expression developed therein models the number of detected counts $\D \N(\en)$ from Eq.~(\ref{dn_detof}), produced by the neutrons of true energy~$E_n$. Here we extend this model on the basis of Eq.~(\ref{dn_detof}), so as to take the resolution function into account.

Let~$\Pij$ denote a particular $\Delta E$-$E$ pair of silicon strips. As in Eq.~(\ref{dn_detof}), let $\D N_\Pij(\etof)$ denote the number of observed counts, detected with the reconstructed energy~$\etof$ by a particular silicon pair~$\Pij$. Further, let~$\C$ represent the relevant carbon isotopes: \mbox{$\C\in\{^{12}\text{C},^{13}\text{C}\}$}. The expression for a differential number of detected counts then takes the form:
\begin{linenomath}\begin{equation}
\frac{\D N_\Pij(\etof)}{\D\etof}=\sum_\C \int_0^\infty \D\en \big[K_{\Pij,\C}(\etof,\en)\sigma_\C(\en)\big].
\label{master}
\end{equation}\end{linenomath}
The sought cross sections~$\sigma_\C(\en)$ for each carbon isotope are here considered as functions of the true neutron energy \textit{in the laboratory frame}. Integration kernels $K_{\Pij,\C}(\etof,\en)$ must be identified separately for each silicon pair~$\Pij$ and carbon isotope~$\C$.

Equation~(\ref{master}) holds separately for the $^\car$C(\textit{n,p}) and for the $^\car$C(\textit{n,d}) reaction. Therefore, all quantities $N_\Pij$, $\sigma_\C$ and $K_{\Pij,\C}$ must be considered separately for each type of reaction, so that the identical but separate analyses can be applied to the $^\car$C(\textit{n,p}) and $^\car$C(\textit{n,d}) reaction.

A form of integration kernels  $K_{\Pij,\C}(\etof,\en)$ was already modeled in~\cite{angular}; here we extend it only by the inclusion of the resolution function $R(\etof,\en)$. For a thin sample (i.e. for \mbox{$\eta_\text{tot}\Sigma_\text{tot}(\en) \ll 1$}; $\eta_\text{tot}$ being the total areal density of the sample and $\Sigma_\text{tot}(E_n)$ the total cross section for \textit{any} reaction to occur, including the neutron elastic scattering), $K_{\Pij,\C}$ terms are:
\begin{linenomath}\begin{align}
\begin{split}
K_{\Pij,\C}(\etof,\en)=\eta_\C R(\etof,\en)  \phi(\en)\sum_{\x=0}^{\X_\C(\en)} \rho_\C(\x,\en) \times &\\
\int_{-1}^1 \D(\cos\theta) \left[\varepsilon_{\Pij,\C}(\x,\en,\cos\theta)  A_\C(\x,\en,\cos\theta)\right]&.
\end{split}
\label{kernel}
\end{align}\end{linenomath}
Here $\eta_\C$ is the areal density (in number of atoms per unit area) of a particular carbon isotope, while $\phi(\en)$ is the neutron flux incident upon the sample, integrated over the duration of measurement.

Neutrons traverse a carbon thickness of 0.35~mm (due to 0.25~mm thin sample being tilted by 45$^\circ$ relative to the beam axis; see Section~\ref{setup}), corresponding to a total areal density of \mbox{$\eta_\text{tot}=4\times10^{-3}$}~atoms/barn. Within the energy range of interest ($10\;\text{MeV}<\en<30\;\text{MeV}$) the total cross section for the natural carbon is \mbox{$\Sigma_\text{tot}(\en) <1.6$~barn} (ENDF/B-VIII.1~\cite{endf}). Thus the thin sample treatment is certainty justified, since \mbox{$\eta_\text{tot}\Sigma_\text{tot}(\en) \lesssim 6\times10^{-3}$}. From $\eta_\text{tot}$ and natural abundances of carbon isotopes, specific isotope densities~$\eta_\C$ are easily determined as \mbox{$\eta_{12}=0.989\eta_\text{tot}$} and \mbox{$\eta_{13}=0.011\eta_\text{tot}$}.

Special care must be taken to account for the possibility of different final states in the daughter nuclei $^{11}$B, $^{12}$B, $^{13}$B. In Eq.~(\ref{kernel}) these states are indicated by~$\x$, ranging from the ground state ($\x=0$) up to the highest state $\X_\C(\en)$ in a specific reaction on a carbon isotope~$\C$. Neutron energy dependent branching ratios \mbox{$\rho_\C(\x,\en)$} govern a probability for a given reaction to proceed through a specific excited state of a daughter nucleus. Angular distributions \mbox{$A_\C(\x,\en,\cos\theta)$} of the reaction products give the probability that a particle is emitted at an angle~$\theta$ relative to the beam direction. Branching ratios are normalized such that:
\begin{linenomath}\begin{equation}
\textstyle \sum_{\x=0}^{\X_\C(\en)} \rho_\C(\x,\en)=1,
\label{br_norm}
\end{equation}\end{linenomath}
and angular distributions are normalized such that:
\begin{linenomath}\begin{equation}
\textstyle \int_{-1}^1 A_\C(\x,\en,\cos\theta) \D(\cos\theta)=1.
\label{ang_norm}
\end{equation}\end{linenomath}
These factors appear in a decomposition of the partial differential cross sections \mbox{$\varrho_\C(\x,\en,\cos\theta)$}:
\begin{linenomath}\begin{equation}
\varrho_\C(\x,\en,\cos\theta)=\sigma_\C(\en) \rho_\C(\x,\en)A_\C(\x,\en,\cos\theta)
\label{part_diff}
\end{equation}\end{linenomath}
for a completely specified reaction outcome. It was shown in~\cite{angular} that the data from our measurement are not sufficient for the separation of angular distributions and branching ratios. Therefore, our goal will be a reconstruction of the angle-integrated and all-states-inclusive cross sections~$\sigma_\C(\en)$, which appears as the main quantity of interest within Eq.~(\ref{master}).

Finally, \mbox{$\varepsilon_{\Pij,\C}(\x,\en,\cos\theta)$} terms from Eq.~(\ref{kernel}) represent the efficiencies for detecting a particular reaction product by a given pair of silicon strips. Since the kinematic parameters (emission energy and the angle-energy correlations) of the emitted products depend on the incident neutron energy and the particular state that the daughter nucleus was left in, the detection deficiencies also inherit these dependences. These efficiencies are determined from the dedicated Geant4 simulations, which are described in detail in~\cite{angular}.

Neutron flux~$\phi(\en)$ in EAR1, appearing in Eq.~(\ref{kernel}), is a well characterized feature of the n\_TOF facility. In the energy range of interest (\mbox{$\en>10$~MeV}) it is determined by the dedicated measurements employing the Parallel Plate Avalanche Counters equipped with the $^{235}$U converter, relying on the $^{235}$U(\textit{n,f}) reaction \cite{flux_ear1}. Within the energy range relevant to this work the evaluated flux shows a smooth $\en$~dependence. For simplicity the \textit{total} flux delivered on the sample has been parameterized by a third degree polynomial:
\begin{linenomath}\begin{equation}
\phi(\en)=\frac{\D \Phi(\en)}{\D\en}=p_0\left(a_0+a_1 E_n +a_2 E_n^2+a_3 E_n^3\right),
\label{flux}
\end{equation}\end{linenomath}
with $\D \Phi(\en)$ the total number of incident neutrons of energy~$\en$, already defined in the context of Eq.~(\ref{profile_def}). Within the energy range $12.5\;\text{MeV}<\en<28.5\;\text{MeV}$, sufficient to this work, the obtained parameter values are:
\begin{linenomath}\begin{gather*}
p_0=184\,525,\, a_0=4258\,\text{MeV}^{-1},\, a_1=-357.4\,\text{MeV}^{-2},\\
a_2=13.01\,\text{MeV}^{-3},\, a_3=-0.1684\,\text{MeV}^{-4}.
\end{gather*}\end{linenomath}
In that, $a_0$, $a_1$, $a_2$, $a_3$ parameterize a nominal neutron flux per single pulse of the neutron beam, i.e. per \mbox{$7\times10^{12}$} protons incident upon the spallation target. Hence, $p_0$ is the total number of pulses from the experiment, corresponding to \textit{$1.292\times10^{18}$} incident protons.

\subsection{Linearizing the model}
\label{linearizing}

Equation~(\ref{master}) is a Fredholm integral equation of the first kind. After proper linearizion it may easily -- though with a certain degree of approximation -- be applied to a discrete set of measured data. Due to the counting statistics from the measurement, we have divided the reconstructed neutron energy range into 1~MeV intervals (bins), as shown in Fig.~\ref{fig_nd_all}. Although a separate and independent binning may be used for a true neutron energy~$\en$ during the cross section reconstruction, we will use the same binning as for the $\etof$ variable.

The lowest coincident detection threshold for protons or deuterons is slightly above 14 MeV. Since the number of detected counts around the detection threshold is very low, the first targeted $\en$ bin will be the one centered at 15~MeV. However, in the analysis we also need to include the preceding bin in $\etof$ due to the n\_TOF resolution function (see below).

\begin{figure}[t!]
\centering
\includegraphics[width=1\linewidth]{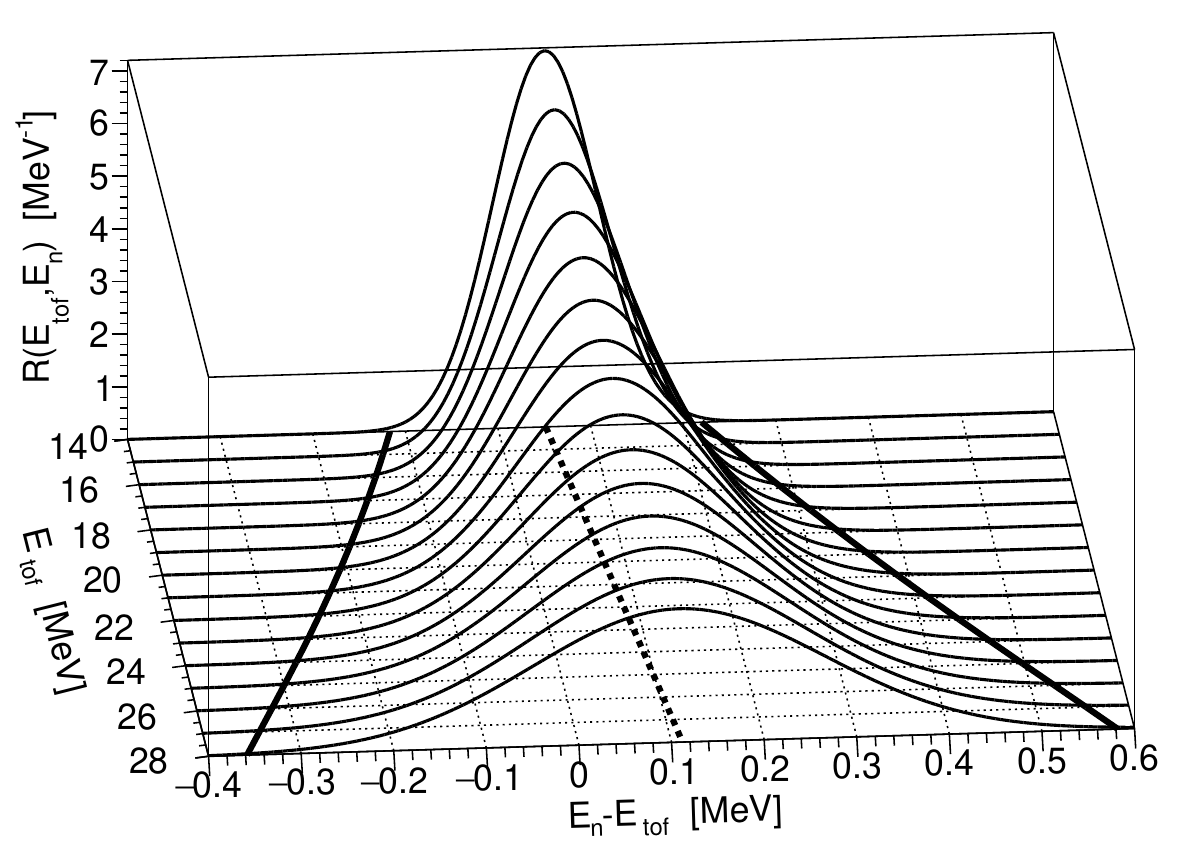}
\caption{Representation of the resolution function from Eq.~(\ref{rf_gaus}), emphasizing $\etof$ dependence of \mbox{$\en-\etof$} deviations. Dashed line follows a mean value of \mbox{$\en-\etof$} distribution at given $\etof$. Solid lines show $\pm3$~standard deviations from the mean value.}
\label{fig_rf}
\end{figure}

Due to the estimated upper bound for the  reliable discrimination of protons and deuterons~\cite{neural}, the last relevant energy bin is the one centered at 25~MeV. In order to assess a range of true neutron energies~$\en$ contributing to the measured counts from the last relevant $\etof$ bin, we again consider the resolution function. Figure~\ref{fig_rf} shows the resolution function parameterized by Eq.~(\ref{rf_gaus}), in a manner appropriate to this analysis. A difference \mbox{$\en-\etof$} is shown, emphasizing the relevant aspects of its variation. Clearly, the relevant part of the resolution function is approximately Gaussian in~$\en$. The thick dashed line shows a mean value of deviations \mbox{$\en-\etof$} for a given~$\etof$. Thick solid lines show a span of $\pm3$~standard deviations from this mean value, indicating a range of true neutron energies $\en$ contributing to the reconstructed energy~$\etof$. At \mbox{$\etof\approx25$~MeV} the upper range of deviations is approximately 0.5~MeV, well within a 1~MeV range of the next energy bin. Therefore, in the subsequent analysis we consider only one energy bin above the targeted range: the one centered at 26~MeV.

Let us denote by index $k$ the reconstructed energy ($\etof$) bins and by index $l$ the true energy ($\en$) bins. Due to the resolution function form (see Fig.~\ref{fig_rf}), only $\en$ bins \mbox{$l=k-1,k,k+1$} may affect the measured counts in a given $\etof$ bin. Integrating Eq.~(\ref{master}) over the $k$-th bin in $\etof$ -- while truncating the contribution to $\en$~integrals from beyond \mbox{$l=k\pm1$} bins -- and linearizing the obtained result so as to decouple the cross section from the integration kernel, yields the following relation for the number of measured counts:
\begin{linenomath}\begin{equation}
N_\Pij^{(k)}\approx  \sum_\C  \sum_{l=k-1}^{k+1}  \left[ \int_k \D\etof \int_l \D\en \:  K_{\Pij,\C}(\etof,\en)  \right] \bar{\sigma}_\C^{(l)},
\label{linear}
\end{equation}\end{linenomath}
with $\bar{\sigma}_\C^{(l)}$ the cross section average over the $l$-th $\en$ bin. For simplicity of expressions, we have denoted by $\int_k$ and $\int_l$ the integrals over appropriate energy bins.

\begin{figure*}[p!] 
\centering
\includegraphics[width=0.5\linewidth]{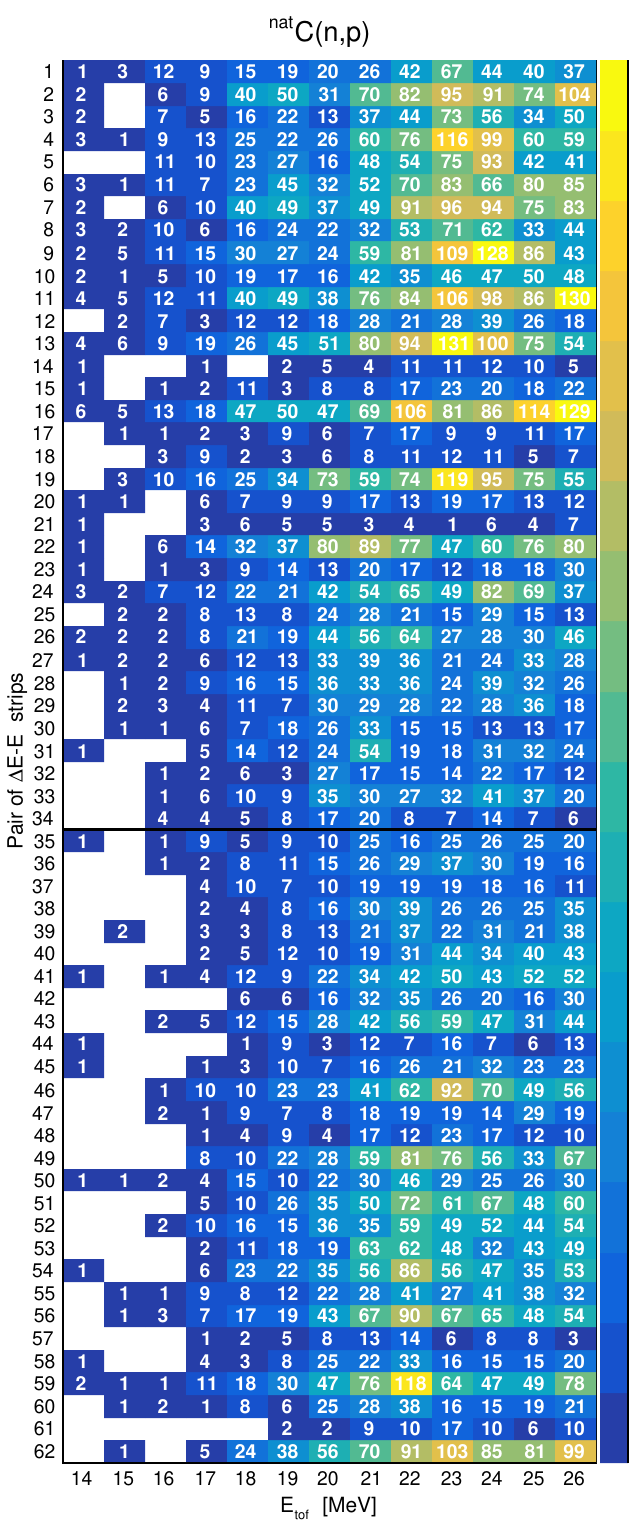}\includegraphics[width=0.5\linewidth]{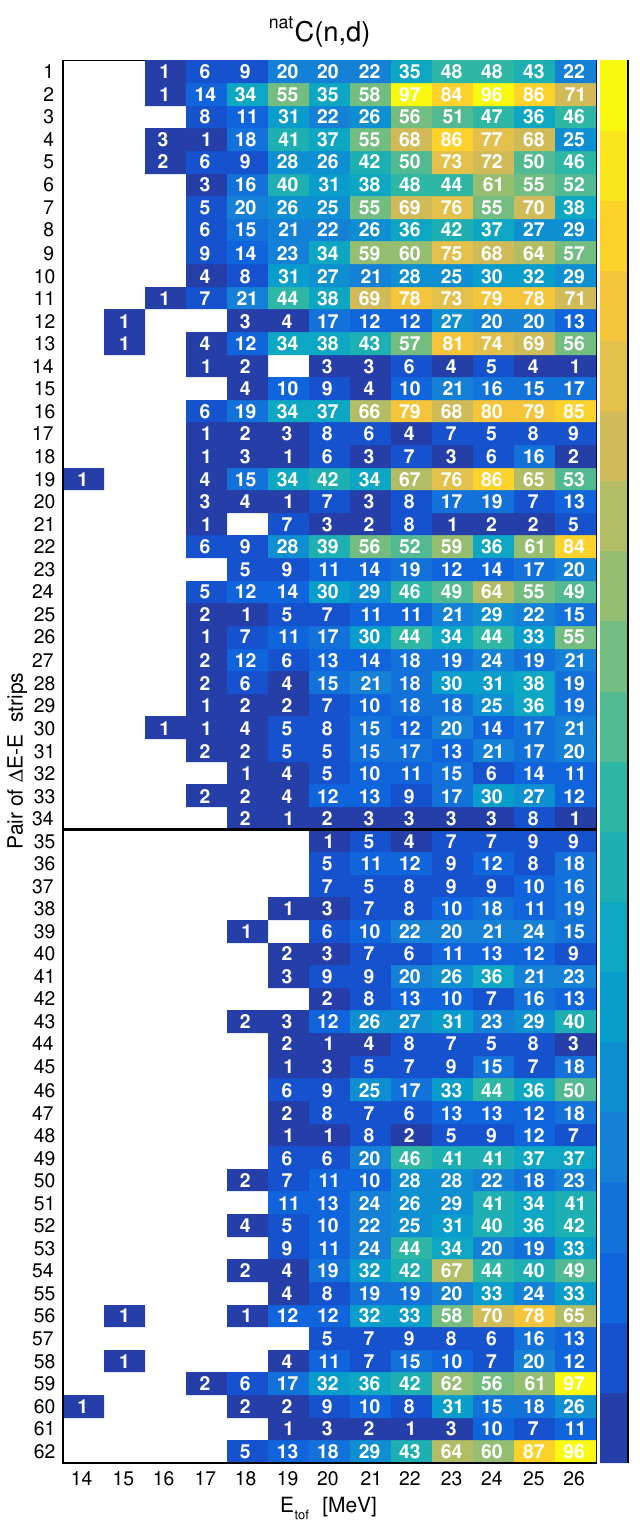}
\caption{Measured number of counts from the $^\car$C(\textit{n,p}) and $^\car$C(\textit{n,d}) reactions, from each relevant pair of silicon strips and per each $\etof$ bin. The content of each cell corresponds to $N_\Pij^{(k)}$ terms from Eq.~(\ref{linear}). The content of the last bin, at 26~MeV, is shown without the correction by the factor $f$ from Eq.~(\ref{f26}). Thick solid line separates the silicon pairs from two silicon telescopes.}
\label{fig_counts}
\end{figure*}

Before solving Eq.~(\ref{linear}) we need to take into account one additional correction. The reconstructed cross section from the last targeted $\en$ bin at 25~MeV will be affected by measured data from the neighboring $\etof$ bin at 26~MeV. However, the $\etof$ data from this bin are also affected by the cross section from the $\en$ bin at 27~MeV. Since we truncate the energy range beyond the 26~MeV bin, we need to account for this surplus of $\etof$ counts in the 26~MeV bin. Assuming the cross section and all relevant factors from Eq.~(\ref{kernel}) to be approximately constant within 25~MeV -- 27~MeV range, the correction factor for the content $N_\Pij^{(k)}$ of the last $\etof$ bin at 26~MeV may be calculated as:
\begin{linenomath}\begin{equation}
f=\frac{\int_{25.5\;\text{MeV}}^{{26.5\;\text{MeV}}}\D\etof \int_{\etof-\mathcal{E}}^{{26.5\;\text{MeV}}} \D\en \times R(\etof,\en)}{\int_{25.5\;\text{MeV}}^{{26.5\;\text{MeV}}}\D\etof \int_{\etof-\mathcal{E}}^{\etof+\mathcal{E}} \D\en \times R(\etof,\en)}=0.88,
\label{f26}
\end{equation}\end{linenomath}
based on the fact that our model considers the $\en$ range only up to 26.5~MeV (thus, a value $f N_\Pij^{(k)}$ is used in the subsequent analysis). In Eq.~(\ref{f26}) the $\en$~width~$\mathcal{E}$ only needs to be large enough for the relevant portion of the resolution function to be covered by the integral. Figure~\ref{fig_rf} clearly indicates that \mbox{$\mathcal{E}=0.6$~MeV} should be sufficient for this purpose.

For both types of reaction -- $^\car$C(\textit{n,p}) and $^\car$C(\textit{n,d}) -- and for each relevant pair~$\Pij$ of silicon strips Fig.~\ref{fig_counts} shows the $N_\Pij^{(k)}$ counts. In the figure the content of the last bin, at 26~MeV, is shown \textit{without} the correction by the factor $f$ from Eq.~(\ref{f26}). Thick solid line separates the silicon pairs from two silicon telescopes, denoted in Fig.~\ref{fig_nd_all} as SITE~1 (\mbox{$\Pij\in[1,34]$}) and SITE~2 (\mbox{$\Pij\in[35,62]$}). The sum of counts from all pairs $\Pij$ in each telescope corresponds to the integral number of counts from Fig.~\ref{fig_nd_all}.

Since the protons and deuterons must have sufficient energy to pass through the $\Delta E$-layer before reaching the $E$-layer, their detection thresholds (around 14.6~MeV for $p_0$ protons and 16.0~MeV for $d_0$ deuterons; see Tab.~\ref{tab_states}) are approximately 1~MeV above their reaction thresholds. Therefore, most of the \mbox{$<$16~MeV} supposed proton counts and \mbox{$<$17~MeV} supposed deuteron counts from Fig.~\ref{fig_counts} are expected to come from the background reactions. They are statistically insignificant and, as such, suppressed by the weighted fitting during the data analysis. For the same reason the final analysis results from Tab.~\ref{tab_results} start at 16~MeV for protons and 17~MeV for deuterons (see the explanation of $\bar{\sigma}_\car$ values in parentheses from Section~\ref{results}).

\subsection{Model adjustments}

We now need to invert Eq.~(\ref{linear}) in order to obtain a set of averaged cross sections $\bar{\sigma}_\C^{(l)}$ from the measured number of counts. Since the system of equations is overdetermined, it can be solved with a least squares minimization approach. We remind here that the analysis requires as the input the angular distributions of emitted charged particles and branching ratios of the excited states of the residual nuclei. As these quantities are not measured directly, assumptions have to be used in the analysis. To this end, different sets of the required input distributions have been used. The main source of such data are the dedicated TALYS-2.0 calculations (Section~\ref{talys_input}), but for purposes of the model-related uncertainty estimation some artificially constructed distributions have also been used (Section~\ref{systematic}). Before going in the details of the input distributions and a further description of the analysis method, some additional adjustments are required.

As a first step, the system from Eq.~(\ref{linear}) can easily be transformed into a clear matrix form. To this end, we introduce two arbitrary bijective mappings:
\begin{linenomath}\begin{equation}
(\Pij,k)\mapsto \alpha \quad\text{and}\quad (\C,l)\mapsto \beta.
\end{equation}\end{linenomath}
Each mapping associates a pair of indices to a \textit{single} index ($\alpha$ or $\beta$). This mapping is basically a ``manual listing'' of one-to-one correspondences between the old indices \mbox{$\Pij,k,\C,l$} and the new ones \mbox{$\alpha,\beta$}, so that from $\alpha(\Pij,k)$ and $\beta(\C,l)$ one can unambiguously recover the original ones as \mbox{$\Pij(\alpha)$}, \mbox{$k(\alpha)$}, \mbox{$\C(\beta)$}, \mbox{$l(\beta)$}. A reduction in the number of indices allows one to express Eq.~(\ref{linear}) in a matrix form:
\begin{linenomath}\begin{equation}
\vec{N}=\mathbf{K}\vec{\sigma},
\label{mx_form}
\end{equation}\end{linenomath}
whose components are:
\begin{linenomath}\begin{align}
&\big[\vec{N}\big]_\alpha=N_\Pij^{(k)},\\
&\big[\vec{\sigma}\big]_\beta = \bar{\sigma}_\C^{(l)},\\
&\big[\mathbf{K}\big]_{\alpha\beta}= \int_{k} \D\etof \int_{l} \D \en \: K_{\Pij,\C}(\etof,\en)\equiv \kappa_{\Pij,\C}^{(k,l)}.
\label{K_matrix}
\end{align}\end{linenomath}
For later clarity in using the original indices \mbox{$\Pij,k,\C,l$}, in Eq.~(\ref{K_matrix}) we have introduced \mbox{$\kappa_{\Pij,\C}^{(k,l)}\equiv\big[\mathbf{K}\big]_{\alpha(\Pij,k)\beta(\C,l)}$} as an equivalent notation for the matrix elements. The $N_\Pij^{(k)}$ counts from Eq.~(\ref{linear}) can now be expressed as:
\begin{linenomath}\begin{equation}
N_\Pij^{(k)}=\sum_{l=k-1}^{k+1} \left( \kappa_{\Pij,12}^{(k,l)} \bar{\sigma}_{12}^{(l)} + \kappa_{\Pij,13}^{(k,l)} \bar{\sigma}_{13}^{(l)} \right).
\label{writeout1}
\end{equation}\end{linenomath}

Let \mbox{$\mathbf{V}=\mathrm{diag}[N_1,N_2,\ldots]$} be a (diagonal) covariance matrix of the experimental counts from~$\vec{N}$. The least-squares solution to an overdetermined system from Eq.~(\ref{mx_form}) is obtained by minimizing a weighted sum of residuals \mbox{$(\vec{N}-\mathbf{K}\vec{\sigma})^\T \mathbf{V}^{-1} (\vec{N}-\mathbf{K}\vec{\sigma})$}, yielding \mbox{$\vec{\sigma}=(\mathbf{K}^\T\mathbf{V}^{-1}\mathbf{K})^{-1} \mathbf{K}^\T\mathbf{V}^{-1}\vec{N}$}~\cite{data_analysis}, wherein $(\cdot)^{-1}$ and $(\cdot)^{\T}$ denote the matrix inverse and transpose, respectively. Applying this method to Eq.~(\ref{mx_form}) recovers, in principle, the cross sections $\bar{\sigma}_{12}^{(l)}$ and $\bar{\sigma}_{13}^{(l)}$ for two carbon isotopes separately. However, the measured number of counts and their statistical uncertainties do not allow for a meaningful separation. In particular, the obtained values $\bar{\sigma}_{13}^{(l)}$ for $^{13}$C feature huge uncertainties. This is because the cross section \mbox{$\sigma_\car=\sum_\C \eta_\C \sigma_\C/\sum_\C \eta_\C$} for a particular reaction on natural carbon is only marginally affected by $^{13}$C isotope due to its low natural abundance of 1.1\% (this is also conditional upon $\sigma_{13}$ not being too large, which is demonstrated in Section~\ref{systematic_talys}). As a consequence, any small (statistical or systematic) variation in the input data from~$\vec{N}$ is easily translated into a large variation in~$\sigma_{13}$. Therefore, from this point on our goal is to reconstruct the cross section values for the natural carbon:
\begin{linenomath}\begin{equation}
\bar{\sigma}_\car^{(l)}\equiv \frac{\eta_{12} \bar{\sigma}_{12}^{(l)} + \eta_{13} \bar{\sigma}_{13}^{(l)}}{\eta_{12}+\eta_{13}}=0.989\bar{\sigma}_{12}^{(l)}+0.011\bar{\sigma}_{13}^{(l)}.
\label{sigma_nat}
\end{equation}\end{linenomath}
Halving the number of extracted parameters will further stabilize the numerical solution for sought $\bar{\sigma}_\car^{(l)}$.

If for each~$\Pij$, $k$ and \mbox{$l\in\{k-1,k,k+1\}$}:
\begin{linenomath}\begin{equation}
\frac{\kappa_{\Pij,\C}^{(k,l)}}{\kappa_{\Pij,12}^{(k,l)} + \kappa_{\Pij,13}^{(k,l)}} = \frac{\eta_\C}{\eta_{12}+\eta_{13}}=
\left\{ \begin{array}{ll} 
0.989 & \text{for}\; \C=12,\\
0.011 & \text{for}\; \C=13,
\end{array} \right.
\label{kappa_eta}
\end{equation}\end{linenomath}
then Eq.~(\ref{writeout1}) becomes:
\begin{linenomath}\begin{equation}
N_\Pij^{(k)}=\sum_{l=k-1}^{k+1} \left( \kappa_{\Pij,12}^{(k,l)} + \kappa_{\Pij,13}^{(k,l)} \right) \bar{\sigma}_\car^{(l)}.
\label{writeout2}
\end{equation}\end{linenomath}
In practice, the equality from Eq.~(\ref{kappa_eta}) is not exact, so that the previous relation becomes approximate. We will show below (see Fig.~\ref{fig_kappa}) that Eq.~(\ref{kappa_eta}) is sufficiently satisfied for an adoption of Eq.~(\ref{writeout2}) to be justified.

In a complete analogy with Eq.~(\ref{mx_form}), the new system of Eqs.~(\ref{writeout2}) may now be represented in a matrix form as:
\begin{linenomath}\begin{equation}
\vec{N}=\mathbf{K}_\car\vec{\sigma}_\car,
\label{master_nat}
\end{equation}\end{linenomath}
with:
\begin{linenomath}\begin{equation}
\big[\mathbf{K}_\car\big]_{\alpha(\Pij,k)\beta(l)}=\kappa_{\Pij,12}^{(k,l)}+\kappa_{\Pij,13}^{(k,l)}.
\label{K_nat}
\end{equation}\end{linenomath}
Due to a loss of sensitivity to a particular carbon isotope, the index~$\beta$ is now determined solely by~$l$ (e.g. \mbox{$\beta=l$}). The overdetermined system from Eq.~(\ref{master_nat}) can now be solved by minimizing a weighted sum of residuals:
\begin{linenomath}\begin{equation}
S_0=(\vec{N}-\mathbf{K}_\car\vec{\sigma}_\car)^\T \mathbf{V}^{-1} (\vec{N}-\mathbf{K}_\car\vec{\sigma}_\car),
\label{residuals_0}
\end{equation}\end{linenomath}
yielding~\cite{data_analysis}: \mbox{$\vec{\sigma}_\car=(\mathbf{K}_\car^\T\mathbf{V}^{-1}\mathbf{K}_\car)^{-1} \mathbf{K}_\car^\T\mathbf{V}^{-1}\vec{N}$}. Reduction in the number of fitted parameters in Eq.~(\ref{writeout2}), relative to Eq.~(\ref{writeout1}), indeed reduces the fluctuations and statistical uncertainties in the calculated cross sections.

One important effect still remains to be taken into account. Let \mbox{$\vec{\mathcal{N}}=\mathbf{K}_\car\vec{\sigma}_\car$} be a vector of counts reconstructed from the optimal solution~$\vec{\sigma}_\car$. These reconstructed counts minimize a weighted global  deviation from the experimental counts. One would naturally expect that the total amount \mbox{$\sum_\alpha \big[\vec{\mathcal{N}}\big]_\alpha$} of reconstructed counts -- for all neutron energies and all pairs of silicon strips -- agrees well with the total amount \mbox{$\sum_\alpha \big[\vec{N}\big]_\alpha$} of experimental counts, thus demonstrating a successful reconstruction of the \textit{integral} cross section. However, independently of the particular choices in the analysis procedure -- whether extracting $\sigma_{12}$ and $\sigma_{13}$ separately, or extracting $\sigma_\car$ directly; whether using TALYS distributions or an artificial set of input distributions (Sections~\ref{talys_input}--\ref{systematic}) -- the integral number of reconstructed counts is somewhat underestimated, implying the underestimation of the (integral) cross section. For this reason it become necessary to impose the requirement of ``conservation'' of the total number of counts during the cross section extraction. This constraint may be implemented by means of Lagrange multiplier(s). The ``conservation'' of counts can be expressed in the matrix form:
\begin{linenomath}\begin{equation}
\textstyle \sum_\alpha \big[\vec{\mathcal{N}}\big]_\alpha=\sum_\alpha \big[\vec{N}\big]_\alpha \quad\Leftrightarrow\quad \vec{\mathrm{I}}^{\,\T}\mathbf{K}_\car\vec{\sigma}_\car=\vec{\mathrm{I}}^{\,\T}\vec{N}
\end{equation}\end{linenomath}
by using a row-vector $\vec{\mathrm{I}}^{\,\T}=[1,1,\ldots,1]$ completely filled with units, having the same dimension as $\vec{N}$. Thus constrained optimization is performed by minimizing:
\begin{linenomath}\begin{equation}
S=S_0+2\lambda \vec{\mathrm{I}}^{\,\T}(\vec{N}-\mathbf{K}_\car\vec{\sigma}_\car),
\label{lagrange_S}
\end{equation}\end{linenomath}
with~$\lambda$ as the Lagrange multiplier. A factor~2 is introduced for convenience. A minimization of~Eq.~(\ref{lagrange_S}) yields a new system of equations that can be written in the extended block-matrix form:
\begin{linenomath}\begin{equation}
\left[ \begin{array}{cc} 
\mathbf{K}_\car^\T\mathbf{V}^{-1}\mathbf{K}_\car & \mathbf{K}_\car^\T\vec{\mathrm{I}} \\
\vec{\mathrm{I}}^{\,\T}\mathbf{K}_\car & 0
\end{array} \right]
\left[ \begin{array}{c} 
\vec{\sigma}_\car \\
\lambda
\end{array} \right]
=
\left[ \begin{array}{c} 
\mathbf{K}_\car^\T\mathbf{V}^{-1} \\
\vec{\mathrm{I}}^{\,\T}
\end{array} \right]\vec{N}.
\label{final_eqs}
\end{equation}\end{linenomath}
Compactly represented, the system takes a form \mbox{$\mathbb{K}\vec{\Sigma}=\mathbb{J}\vec{N}$}, with the column-vector \mbox{$\vec{\Sigma}=[\vec{\sigma}_\car,\lambda]^\T$} containing the sought cross section values from~$\vec{\sigma}_\car$. With \mbox{$\mathbb{L}\equiv (\mathbb{K}^\T\mathbb{K})^{-1}\mathbb{K}^\T\mathbb{J}$}, the formal solution to this optimization problem is easily expressed as $\vec{\Sigma}=\mathbb{L}\vec{N}$, together with the covariance matrix of the extracted results \mbox{$\mathbf{V}_{\vec{\Sigma}}=\mathbb{L} \mathbf{V}\mathbb{L}^\T$}.

\subsection{Model inputs: TALYS data}
\label{talys_input}

The final step of the analysis is to construct the matrix~$\mathbf{K}_\car$ -- defined by Eqs.~(\ref{kernel}), (\ref{K_matrix}) and~(\ref{K_nat}) -- by adopting a particular set of branching ratios \mbox{$\rho_\C(\x,\en)$} and angular distributions \mbox{$A_\C(\x,\en,\cos\theta)$} from some external source of information. For this purpose we use the nuclear reaction code TALYS-2.0~\cite{talys_2012,talys_2023,talys_web}.

TALYS code contains several macroscopic and (semi)microscopic models for each physical quantity used in calculations. The selection of a model is given by the values of individual parameters that correspond to each quantity in TALYS, see Tab.~\ref{tab_models}. Unfortunately, it is not clear which of the parameter values best describe a specific nucleus. This selection can strongly influence the calculated quantities. As an illustration, an impact of different parameter values on the calculated cross sections is demonstrated in the TENDL~\cite{tendl} and the TENDL-astro library~\cite{tendl_astro_web,tendl_astro}.

\begin{table}[t!]
\caption{TALYS model parameters, their basic description and values used in this work. Not all combinations of listed values were considered. The values that were considered only in combination with some other specific value are denoted by shared superscript (a, b, c, d, e), making a total of 480 model sets. For example, \mbox{strength=8} was combined only with \mbox{strengthm1=8} (as denoted by a superscript~'a').} 
\centering

\begin{tabular}{ccccc}
\hline\hline
\textbf{Parameter} &\hrz& \textbf{Meaning} &\hrz& \textbf{Values}\\
\hline
strength && gamma strength function for E1 && 8$^\text{a}$, 9$^\text{b}$ \\
ldmodel && level density && 1$^\text{c}$, 2$^\text{d}$, 5$^\text{e}$ \\
jlmomp && optical model && n, y \\
strengthm1 && gamma strength function for M1 && 3$^\text{b}$, 8$^\text{a}$ \\
colenhance && collective enhancement && n$^\text{cde}$, y$^\text{cd}$ \\
widthmode && width fluctuation && 0, 1, 2 \\
massmodel && mass model && 0, 1, 2, 3 \\
alphaomp && alpha optical model  && 5, 6 \\
fismodel && fission model && 1 \\
\hline\hline
\end{tabular}

\label{tab_models}
\end{table}

We decided to use several sets of parameter values. Each of them will be referred as a \textit{model set}. Using several sets allows us to produce variations in the calculated cross sections for individual final levels (from which we extract the required branching ratios and angular distributions) and to investigate the sensitivity of our analysis results to these variations. Table~\ref{tab_models} lists a set of 9 available model parameters, their basic description and the parameter values considered in this work. A more detailed description of their meaning and available values may be found in \reffs~\cite{tendl_astro_web,tendl_astro,talys_keff,talys_web}. Not all combinations of the listed parameter values were investigated, which is indicated in Tab.~\ref{tab_models} by the shared superscripts. The parameter values without a superscript were combined with all other listed values. Those with a given superscript were combined only with the parameter values sharing the same superscript, making a total of 480 considered model sets. Our final analysis results are based on 384 of these model sets (see Section~\ref{systematic_talys}).

\newcommand{\colA}{ $\boldsymbol{\x}$ }
\newcommand{\colB}{ $\boldsymbol{E_\x}$ }
\newcommand{\colC}{ $\boldsymbol{Q_\x}$ }
\newcommand{\colD}{ $\boldsymbol{E_\mathrm{thr}^{(\x)}}$ }
\newcommand{\colE}{ $\boldsymbol{E_\mathrm{det}^{(\x)}}$ }

\begin{table*}[t!]
\caption{Excited states (plus ground state) for four relevant reactions, from three relevant residual boron nuclei, considered in the analysis of the $^\car$C(\textit{n,p}) and $^\car$C(\textit{n,d}) reactions. All energies are expressed in~MeV. See the main text for the explanation of all listed quantities.} 
\centering

\begin{tabular}{ccc|ccccccccc|ccccccccc|ccccccccc|ccccccccc}
\hline\hline

&&& \multicolumn{9}{c|}{\vrt\textbf{$\boldsymbol{^{12}}$C(\textit{n,p})$\boldsymbol{^{12}}$B}\vrt} & 
\multicolumn{9}{c|}{\vrt\textbf{$\boldsymbol{^{13}}$C(\textit{n,p})$\boldsymbol{^{13}}$B}\vrt} & 
\multicolumn{9}{c|}{\vrt\textbf{$\boldsymbol{^{12}}$C(\textit{n,d})$\boldsymbol{^{11}}$B}\vrt} & 
\multicolumn{9}{c}{\vrt\textbf{$\boldsymbol{^{13}}$C(\textit{n,d})$\boldsymbol{^{12}}$B}\vrt} \\
\hline
\hrz& \colA &\hrz&\hrz& \colB && \colC && \colD && \colE &\hrz&\hrz& \colB && \colC && \colD && \colE &\hrz&\hrz& \colB && \colC && \colD && \colE &\hrz&\hrz& \colB && \colC && \colD && \colE &\hrz \\
\hline
&0   &&& 0.00 && 12.59 && 13.65 && 14.6 &&& 0.00 && 12.65 && 13.64 && 14.6 &&& 0.00 && 13.73 && 14.90 && 16.0 &&& 0.00 && 15.31 && 16.51 && 17.6 &\\
&1   &&& 0.95 && 13.54 && 14.69 && 15.7 &&& 3.48 && 16.14 && 17.40 && 18.3 &&& 2.12 && 15.86 && 17.20 && 18.2 &&& 0.95 && 16.26 && 17.53 && 18.7 &\\
&2   &&& 1.67 && 14.26 && 15.47 && 16.4 &&& 3.53 && 16.19 && 17.46 && 18.4 &&& 4.44 && 18.18 && 19.72 && 20.6 &&& 1.67 && 16.98 && 18.31 && 19.4 &\\
&3   &&& 2.62 && 15.21 && 16.50 && 17.4 &&& 3.68 && 16.33 && 17.61 && 18.6 &&& 5.02 && 18.75 && 20.35 && 21.3 &&& 2.62 && 17.93 && 19.33 && 20.3 &\\
&4   &&& 2.72 && 15.31 && 16.61 && 17.5 &&& 3.71 && 16.37 && 17.65 && 18.6 &&& 6.74 && 20.47 && 22.22 && 23.1 &&& 2.72 && 18.03 && 19.44 && 20.5 &\\
&5   &&& 3.39 && 15.98 && 17.33 && 18.2 &&& 4.13 && 16.78 && 18.10 && 19.0 &&& 6.79 && 20.52 && 22.27 && 23.2 &&& 3.39 && 18.70 && 20.16 && 21.2 &\\
&6   &&& 3.76 && 16.35 && 17.73 && 18.6 &&& 4.83 && 17.48 && 18.85 && 19.8 &&& 7.29 && 21.02 && 22.81 && 23.6 &&& 3.76 && 19.07 && 20.56 && 21.6 &\\
&7   &&& 4.00 && 16.59 && 18.00 && 18.9 &&& 5.02 && 17.68 && 19.06 && 20.0 &&& 7.98 && 21.71 && 23.56 && 24.3 &&& 4.00 && 19.31 && 20.82 && 21.8 &\\
&8   &&& 4.30 && 16.89 && 18.32 && 19.1 &&& 5.11 && 17.76 && 19.15 && 20.1 &&& 8.56 && 22.29 && 24.19 && 24.9 &&& 4.30 && 19.61 && 21.15 && 22.1 &\\
&9   &&& 4.46 && 17.05 && 18.49 && 19.3 &&& 5.39 && 18.04 && 19.46 && 20.3 &&& 8.92 && 22.65 && 24.58 && 25.3 &&& 4.46 && 19.77 && 21.32 && 22.3 &\\
&10 &&& 4.52 && 17.11 && 18.56 && 19.3 &&& 5.56 && 18.21 && 19.64 && 20.6 &&& 9.18 && 22.92 && 24.87 && 25.6 &&& 4.52 && 19.83 && 21.39 && 22.4 &\\
&11 &&& 4.99 && 17.58 && 19.07 && 19.9 &&& 6.17 && 18.82 && 20.30 && 21.2 &&& 9.27 && 23.00 && 24.96 && 25.7 &&& 4.99 && 20.30 && 21.89 && 22.8 &\\
&12 &&& 5.61 && 18.20 && 19.74 && 20.5 &&& 6.42 && 19.08 && 20.57 && 21.5 &&&  &&  &&  &&  &&& 5.61 && 20.92 && 22.56 && 23.5 &\\
&13 &&& 5.73 && 18.31 && 19.87 && 20.6 &&& 6.93 && 19.59 && 21.12 && 22.0 &&&  &&  &&  &&  &&& 5.73 && 21.03 && 22.68 && 23.6 &\\
&14 &&& 6.00 && 18.59 && 20.17 && 21.0 &&& 7.52 && 20.17 && 21.75 && 22.6 &&&  &&  &&  &&  &&& 6.00 && 21.31 && 22.98 && 23.9 &\\
&15 &&& 6.20 && 18.79 && 20.38 && 21.2 &&& 7.86 && 20.51 && 22.12 && 23.0 &&&  &&  &&  &&  &&& 6.20 && 21.51 && 23.20 && 24.1 &\\
&16 &&& 6.60 && 19.19 && 20.82 && 21.6 &&& 8.13 && 20.79 && 22.42 && 23.2 &&&  &&  &&  &&  &&& 6.60 && 21.91 && 23.63 && 24.5 &\\
&17 &&& 7.06 && 19.65 && 21.32 && 22.1 &&& 8.68 && 21.34 && 23.01 && 23.8 &&&  &&  &&  &&  &&& 7.06 && 22.37 && 24.12 && 25.0 &\\
&18 &&& 7.30 && 19.89 && 21.58 && 22.3 &&& 9.44 && 22.09 && 23.83 && 24.6 &&&  &&  &&  &&  &&& 7.30 && 22.61 && 24.38 && 25.2 &\\
&19 &&& 7.54 && 20.13 && 21.84 && 22.6 &&& 9.50 && 22.15 && 23.89 && 24.7 &&&  &&  &&  &&  &&& 7.54 && 22.85 && 24.65 && 25.5 &\\
&20 &&& 7.67 && 20.26 && 21.98 && 22.7 &&& 10.22 && 22.87 && 24.67 && 25.4 &&&  &&  &&  &&  &&& 7.67 && 22.98 && 24.78 && 25.6 &\\
&21 &&& 7.70 && 20.29 && 22.01 && 22.7 &&&  &&  &&  &&  &&&  &&  &&  &&  &&& 7.70 && 23.01 && 24.81 && 25.6 &\\
&22 &&& 7.80 && 20.39 && 22.12 && 22.9 &&&  &&  &&  &&  &&&  &&  &&  &&  &&& 7.80 && 23.11 && 24.92 && 25.7 &\\
&23 &&& 7.84 && 20.42 && 22.16 && 22.9 &&&  &&  &&  &&  &&&  &&  &&  &&  &&& 7.84 && 23.14 && 24.96 && 25.8 &\\
&24 &&& 7.94 && 20.52 && 22.27 && 23.0 &&&  &&  &&  &&  &&&  &&  &&  &&  &&&  &&  &&  &&  &\\
&25 &&& 8.13 && 20.72 && 22.48 && 23.2 &&&  &&  &&  &&  &&&  &&  &&  &&  &&&  &&  &&  &&  &\\
&26 &&& 8.16 && 20.75 && 22.52 && 23.3 &&&  &&  &&  &&  &&&  &&  &&  &&  &&&  &&  &&  &&  &\\
&27 &&& 8.24 && 20.83 && 22.60 && 23.4 &&&  &&  &&  &&  &&&  &&  &&  &&  &&&  &&  &&  &&  &\\
&28 &&& 8.39 && 20.98 && 22.76 && 23.5 &&&  &&  &&  &&  &&&  &&  &&  &&  &&&  &&  &&  &&  &\\
&29 &&& 8.58 && 21.17 && 22.97 && 23.7 &&&  &&  &&  &&  &&&  &&  &&  &&  &&&  &&  &&  &&  &\\
&30 &&& 8.71 && 21.30 && 23.11 && 23.8 &&&  &&  &&  &&  &&&  &&  &&  &&  &&&  &&  &&  &&  &\\

\hline\hline
\end{tabular}

\label{tab_states}
\end{table*}

For each model set TALYS provides the required branching ratios \mbox{$\rho_\C(\x,\en)$} and angular distributions \mbox{$A_\C(\x,\en,\cos\theta)$} to be used in Eq.~(\ref{kernel}), for each excited state~$\x$ of the relevant boron nuclei (for simplicity of terminology, the ground state is also included among~$\x$ as the ``zeroth excited state'' \mbox{$\x=0$}). In particular, TALYS provides the partial differential cross sections \mbox{$\varrho_\C(\x,\en,\cos\theta)$}, given by Eq.~(\ref{part_diff}). From these the angular distributions are obtained by normalizing them according to Eq.~(\ref{ang_norm}), so that:
\begin{linenomath}\begin{equation}
A_\C(\x,\en,\cos\theta)=\frac{\varrho_\C(\x,\en,\cos\theta)}{\int_{-1}^1 \varrho_\C(\x,\en,\cos\theta') \D(\cos\theta')},
\end{equation}\end{linenomath}
while the branching ratios are constructed so as to satisfy the normalization from Eq.~(\ref{br_norm}):
\begin{linenomath}\begin{equation}
\rho_\C(\x,\en)= \frac{\int_{-1}^1 \varrho_\C(\x,\en,\cos\theta) \D(\cos\theta)}{\sum_{\x'=0}^{\X_\C(\en)} \int_{-1}^1 \varrho_\C(\x',\en,\cos\theta) \D(\cos\theta)}.
\end{equation}\end{linenomath}
The list and the properties of the excited states of three relevant boron isotopes -- $^{11}$B~\cite{b11_states}, $^{12}$B~\cite{b12_states}, $^{13}$B~\cite{b13_states} -- are readily available. For each of four relevant reactions Table~\ref{tab_states} lists the excited states considered in this work, according to their excitation energy~$E_\x$. The relevant properties of these states also include the total width, spin and parity assignment. We do not include these properties here; they may be found in~\cite{b11_states,b12_states,b13_states}. Alongside the excitation energy~$E_\x$ we also list the $Q$-values $Q_\x$ for each particular reaction to proceed via given excited state (for simplicity, we treat the $Q$-values as positive):
\begin{linenomath}\begin{equation}
Q_\x=Q_0+E_\x,
\label{q_value}
\end{equation}\end{linenomath}
where~$Q_0$ is the $Q$-value for the ground state of a particular boron isotope. We also list the energy thresholds~$E_\mathrm{thr}^{(\x)}$ in the laboratory frame (a frame of a carbon nucleus~$\C$ at rest), obtained from a relativistic expression:
\begin{linenomath}\begin{equation}
E_\mathrm{thr}^{(\x)}=\left(1+\frac{m_n}{m_\C}\right)Q_\x+\frac{Q_\x^2}{2m_\C c^2}.
\label{e_thr}
\end{equation}\end{linenomath}
Finally, we provide a list of approximate detection thresholds $E_\mathrm{det}^{(\x)}$, read out from the results of Geant4 simulations.

\begin{figure}[t!]
\centering
\includegraphics[width=0.78\linewidth]{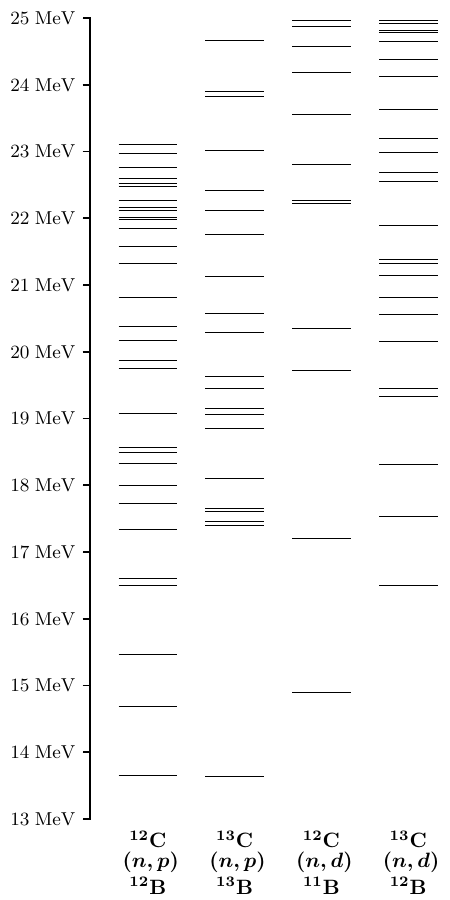}
\caption{Excited states from Tab.~\ref{tab_states}, included in the analysis of the $^\car$C(\textit{n,p}) and $^\car$C(\textit{n,d}) reactions, ordered according to their energy thresholds~$E_\mathrm{thr}^{(\x)}$ in the laboratory frame.}
\label{fig_states}
\end{figure}

Figure~\ref{fig_states} shows all excited states included in our analysis. When possible, we have considered the states with the laboratory thresholds \mbox{$E_\mathrm{thr}^{(\x)}<25$}~MeV. The only exception is the $^{12}$C(\textit{n,p})$^{12}$B reaction, since TALYS does not allow the inclusion of states above the 30th excited one. However, a combination of the low branching ratios close to their reaction thresholds, and of a reduced detection efficiency close to the coincident detection threshold makes a contribution of the excluded states negligible. We have confirmed this by performing two separate analyses with an artificial set of branching ratios and angular distributions (see Section~\ref{systematic}): one taking into account only the states up to the 30th excited one, the other taking into account the higher states. A difference in the extracted cross section for the $^\car$C(\textit{n,p}) reaction amounts to 1.3\% within the extraneous 26~MeV bin (see Section~\ref{linearizing}) and only 0.06\% within the last targeted bin at 25~MeV.

\subsection{Model-related uncertainties: TALYS variations}
\label{systematic_talys}

The model-related uncertainties in the angular distributions and branching ratios generated by TALYS-2.0 have been estimated by using two methods. The first method consists in varying the TALYS model parameters -- i.e. in considering multiple model sets from Tab.~\ref{tab_models} -- and observing the spread of the analysis results. For the $^\car$C(\textit{n,p}) reaction the RMS of the results obtained with different model sets is well below 1\% and is thus negligible. Later Tab.~\ref{tab_results} ($\Delta \sigma_\mathrm{talys}$ column) and Fig.~\ref{fig_vars} show that the difference between model sets becomes notable for the $^\car$C(\textit{n,d}) reaction above 20~MeV. We observe that there is no ``continuous'' spread of the results with parameter variations. Rather, the analysis results form two very narrow groups of values. We find that the difference between these two groups is entirely determined by a single \mbox{``ldmodel''} parameter. In particular, parameter values~1 and~5 produce the lower cross sections, while a value~2 provides the higher cross sections. Value~1 corresponds to a phenomenological constant temperature + Fermi gas model; value~2 to a phenomenological back-shifted Fermi gas model; value~5 to the microscopic Skyrme-Hartree-Fock-Bogolyubov model using combinatorial level densities from numerical tables~\cite{talys_web}. It should be noted that models~1 and~5 are fundamentally different; there is no \textit{a~priori} reason for their results to agree, as they do in this case.

\begin{figure}[t!]
\centering
\includegraphics[width=1\linewidth]{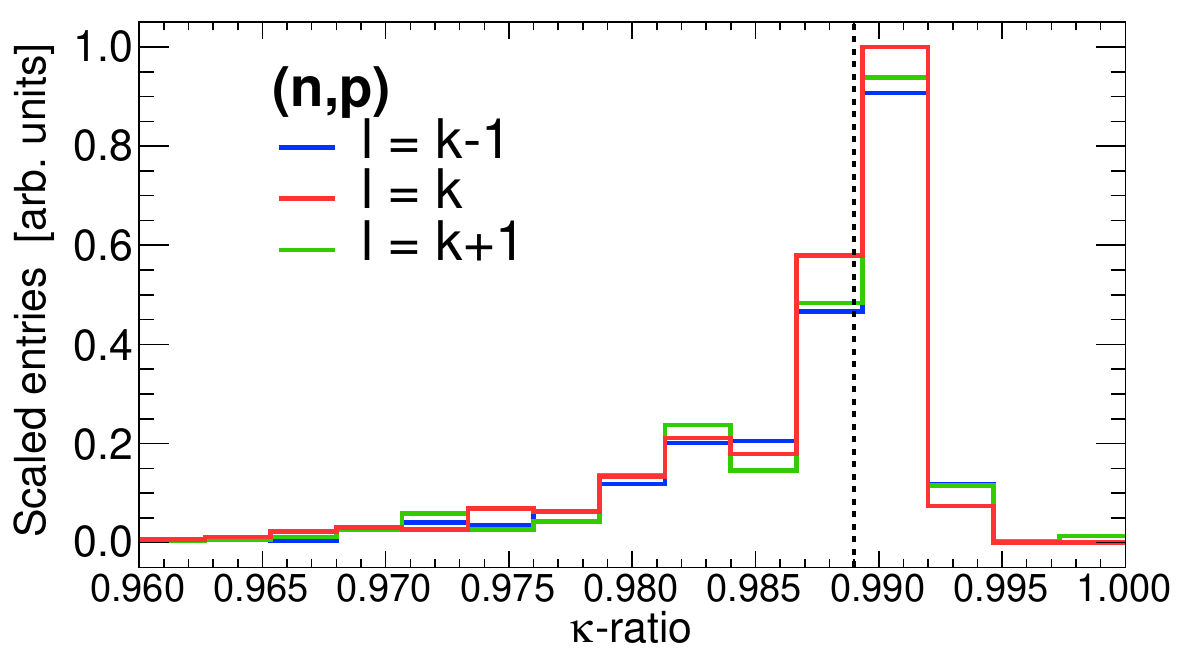}
\includegraphics[width=1\linewidth]{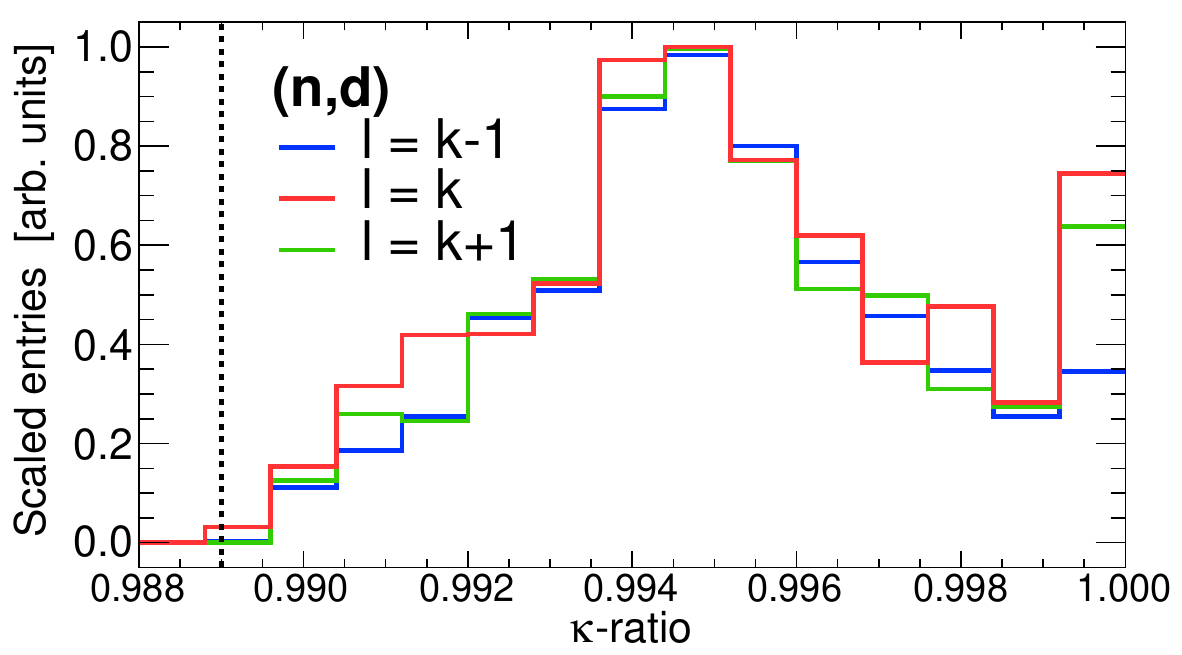}
\caption{Distributions of $\kappa$-ratios from Eq.~(\ref{kappa_eta}) for $^{12}$C, arbitrarily scaled so that the \mbox{$l=k$} distributions peak at 1. Dashed vertical lines mark the $^{12}$C abundance of 0.989.} 
\label{fig_kappa}
\end{figure}

The final cross section results are obtained by averaging the results from separate analyses performed with different model sets. Among~480 models sets indicated by the model combinations from Tab.~\ref{tab_models}, a total of 192~model sets were obtained with \mbox{ldmodel=1}, 192~model sets with \mbox{ldmodel=2} and 96~model sets with \mbox{ldmodel=5}. Therefore, 288~model sets yield the lower (but similar) cross section values, while only 192 of them yield the higher values. Averaging all these model sets indiscriminately would bias the final results towards lower values for no valid reason. In order to provide a conservative and unbiased evaluation of the cross sections, for the final averaging we consider the same number of model sets predicting higher and lower cross section: all those with the ``ldmodel'' values~1 or~2, making a total of 384 model sets. The results obtained in this way are reported in Tab.~\ref{tab_results} as~$\bar{\sigma}_\car$. For the estimation of the model related uncertainties see Sec.~\ref{results}.

Let us now address the validity of a condition from Eq.~(\ref{kappa_eta}). Figure~\ref{fig_kappa} shows a distribution of $\kappa$-ratios \mbox{$\kappa_{\Pij,12}^{(k,l)}/\big(\kappa_{\Pij,12}^{(k,l)} + \kappa_{\Pij,13}^{(k,l)}\big)$} for the (\textit{n,p}) and (\textit{n,d}) reactions, and for each \mbox{$l\in\{k-1,k,k+1\}$} separately. Only the distributions for $^{12}$C (i.e. \mbox{$\C=12$}) are shown, since the distributions for \mbox{$\C=13$} are perfectly mirrored around $^{13}$C abundance of~0.011. The distributions were constructed from the matrix terms for all 62 relevant $\Delta E$-$E$ pairs of silicon strips, and from all 384 model sets used in the final analysis. A mean value of the \mbox{$l=k$} distribution (corresponding to the matrix terms that dominantly affect the final results) is~0.985 for the (\textit{n,p}) reaction and~0.994 for the (\textit{n,d}) reaction, to be compared against the $^{12}$C abundance of~0.989 (shown by the dashed vertical lines). Evidently, the condition from Eq.~(\ref{kappa_eta}) for the validity of Eq.~(\ref{writeout2}) is satisfied to a high degree for both reactions.

\begin{figure}[t!]
\centering
\includegraphics[width=1\linewidth]{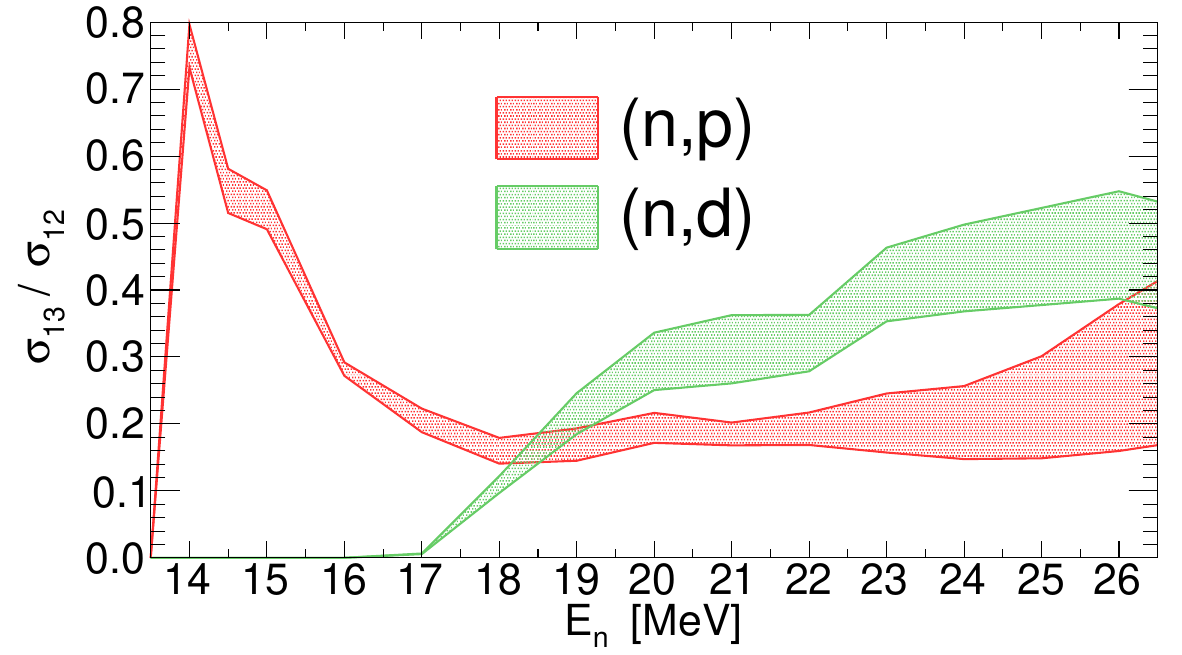}
\caption{Ratio between $^{13}$C and $^{12}$C cross sections from 480 considered TALYS-2.0 model sets. Shaded areas show a range between a minimum and maximum observed ratio.}
\label{fig_ratio_talys}
\end{figure}

\begin{figure}[t!]
\centering
\includegraphics[width=1\linewidth]{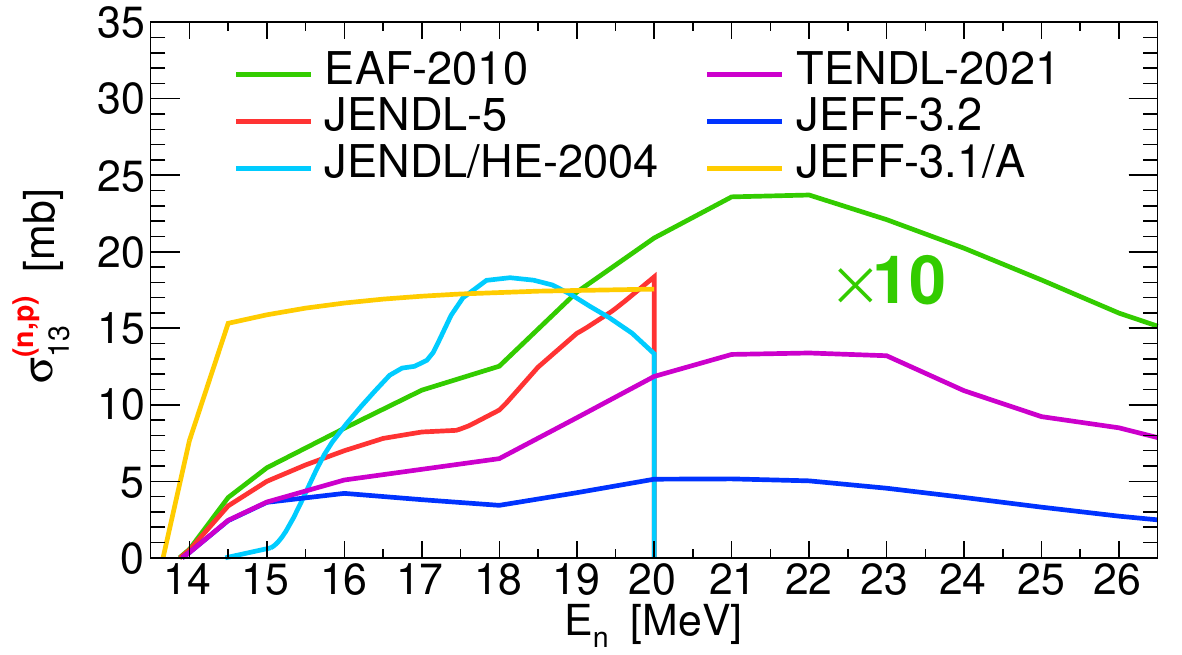}
\includegraphics[width=1\linewidth]{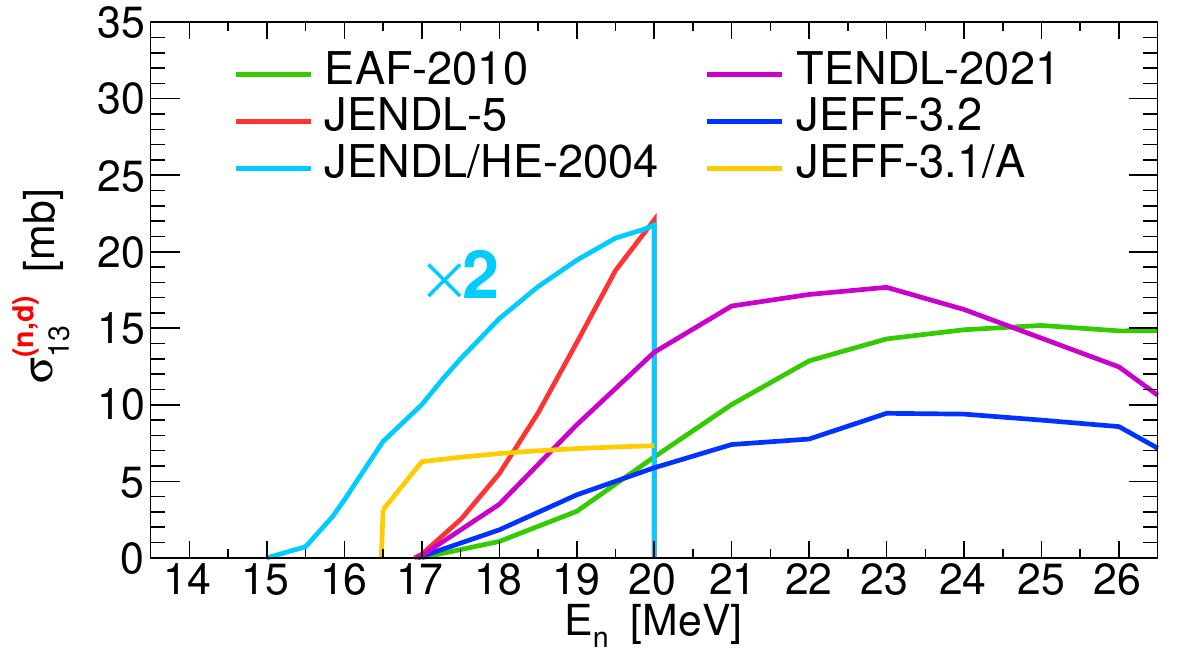}
\caption{Selected $^{13}$C(\textit{n,p}) and $^{13}$C(\textit{n,d}) cross section evaluations. EAF-2010 (\textit{n,p}) cross section is 10 times higher than displayed; JENDL/HE-2004 (\textit{n,d})  cross section is 2 times higher than displayed.}
\label{fig_13c}
\end{figure}

It is also useful to verify that the cross sections for $^{13}$C are not much higher than those for $^{12}$C. Since the validity of Eq.~(\ref{writeout2}) is conditional upon Eq.~(\ref{kappa_eta}), the goal is to ensure that the contribution from~$\sigma_{13}$ is not so large that it starts interfering with the approximate validity of Eq.~(\ref{kappa_eta}). To this end we examine a ratio $\sigma_{13}/\sigma_{12}$ of the cross sections from TALYS calculations. Figure~\ref{fig_ratio_talys} shows the spread of such ratios extracted from all 480 considered model sets. The cross sections for $^{13}$C are predicted to be lower than those from $^{12}$C, implying that $\sigma_\car$ can be safely assumed to be dominated by~$\sigma_{12}$, due to a high natural abundance of $^{12}$C. Since TALYS is not optimally suited for light nuclei, the ratios from Fig.~\ref{fig_ratio_talys} should be considered with some caution. For this reason, Fig.~\ref{fig_13c} shows several cross section evaluations for $^{13}$C, from various evaluation libraries. Most of them peak below 25~mb. The only exceptions are the \mbox{EAF-2010} cross section for the (\textit{n,p}) reaction, which has been rescaled by a factor 0.1 (and is therefore 10 times higher), and the \mbox{JENDL/HE-2004} cross section for the (\textit{n,d}) reaction, which has been rescaled by a factor 0.5 (being twice as high). Compared with $^{12}$C evaluations shown in Fig.~\ref{fig_tendl}, most of $^{13}$C cross sections are below, or at least of the same order of magnitude as the cross sections for $^{12}$C, justifying our approach.

\subsection{Model-related uncertainties: artificial distributions}
\label{systematic}

The other method for estimating the systematic uncertainties related to the calculated angular distributions and branching ratios consists in performing an independent analysis using an artificial and deliberately simplistic set of branching ratios and angular distributions. We consider such distributions as representing a significant deviation from any realistic set. Accordingly, a difference between the analysis results obtained with TALYS and those obtained with artificial distributions should provide a reasonable and conservative estimate of the related systematic effects.

Isotropic angular distribution \mbox{$A_\C(\x,\en,\cos\theta)=1/2$} is a natural candidate for both the (\textit{n,p}) and (\textit{n,d}) reaction, for both carbon isotopes~$\C$, for each excited state~$\x$ and for every value of neutron energy~$\en$.

\begin{figure}[t!]
\centering
\includegraphics[width=1\linewidth]{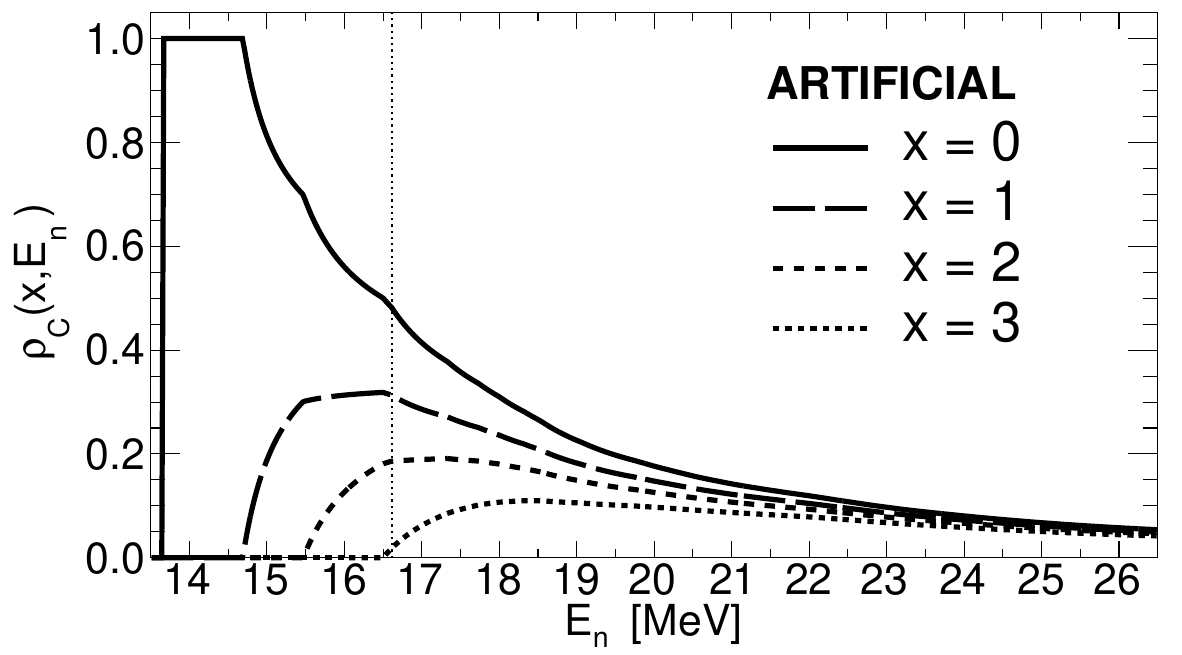}
\includegraphics[width=1\linewidth]{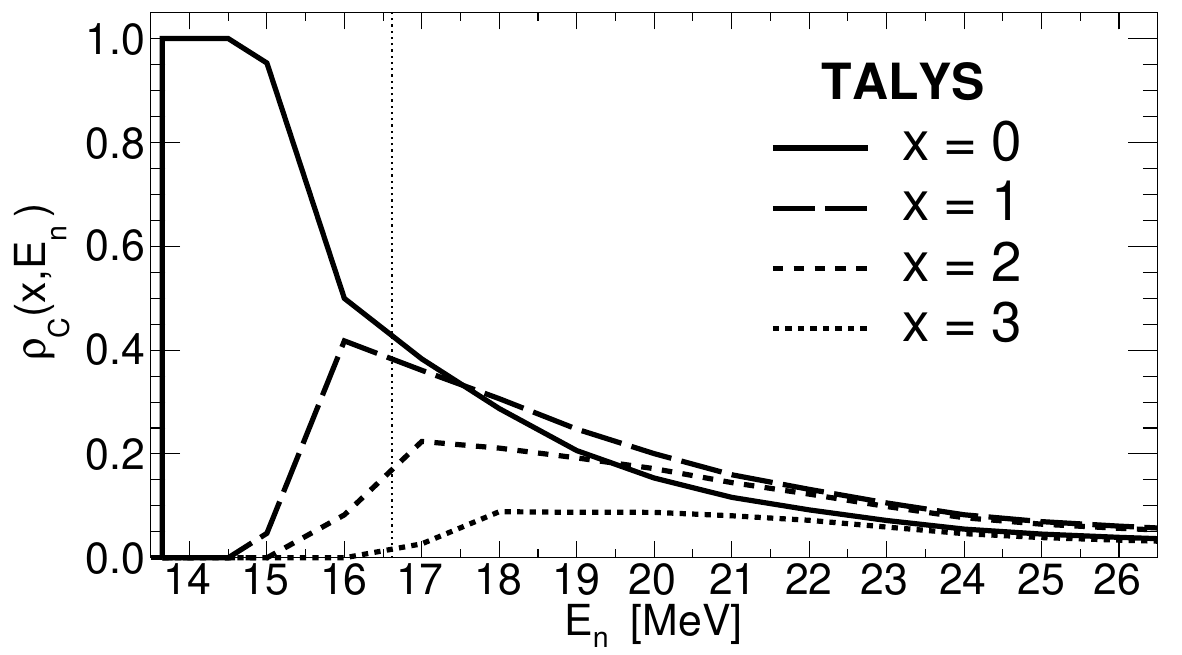}
\vspace*{-4mm}
\caption{Branching ratios for the $^{12}$C(\textit{n,p})$^{12}$B reaction, for the first four states in $^{12}$B. Top: artificial constructions from Eq.~(\ref{br_art}). Bottom: branching ratios from TALYS, from an arbitrarily selected model set. Beyond the vertical line at 16.61~MeV the sum of shown contributions is not equal to~1 due to the presence of higher excited states, not shown here.}
\label{fig_branch}
\vspace*{-2mm}
\end{figure}

In order to construct artificial branching ratios we first introduce a following simplistic function:
\begin{linenomath}\begin{equation}
f(x)=\left\{\begin{array}{ccc}
x & \text{if} & x>0,\\
0 & \text{if} & x\le0,
\end{array}\right. 
\end{equation}\end{linenomath}
and use it in constructing the artificial set:
\begin{linenomath}\begin{equation}
\rho_\C(\x,\en)=\frac{f\big[\mathcal{E}_\C(\en)-Q_\x\big]}{\sum_{\x'}f\big[\mathcal{E}_\C(\en)-Q_{\x'}\big]}.
\label{br_art}
\end{equation}\end{linenomath}
For the neutron energy~$\en$ in the laboratory frame, the term $\mathcal{E}_\C(\en)$:
\begin{linenomath}\begin{equation}
\mathcal{E}_\C(\en)=\left[\sqrt{1+\frac{2 m_\C \en}{(m_n+m_\C)^2c^2}}-1\right] \! (m_n+m_\C)c^2
\end{equation}\end{linenomath}
corresponds to a total kinetic energy in the center of mass frame \textit{before} a particular reaction -- (\textit{n,p}) or (\textit{n,d}) -- on a particular carbon isotope~$\C$. Thus, the terms in the square brackets in Eq.~(\ref{br_art}) represent a total kinetic energy in the center of mass frame \textit{after} the reaction proceeds via any of the excited states. A construction from Eq.~(\ref{br_art}) ensures that the branching ratio for a given excited state vanishes below its energy threshold; that it starts smoothly from 0 at the reaction threshold and peaks close to it; and that it trails off at higher energies, becoming asymptotically equal to all other branching ratios.

Figure~\ref{fig_branch} shows an example of the artificially constructed branching ratios for the $^{12}$C(\textit{n,p})$^{12}$B reaction, for the first four states in $^{12}$B. They are compared with the branching ratios from a single, arbitrarily selected TALYS model set. Above the threshold for the 4th excited state (vertical line at 16.61~MeV; see Table~\ref{tab_states}), the sum of displayed branching ratios is no longer equal to~1 due to the presence of higher excited states not shown in the figure.

\vspace*{-3mm}

\section{Results and discussion}
\label{results}

\begin{table*}[t!]
\caption{Energy-differential cross section for the $^{\car}$C(\textit{n,p}) and $^{\car}$C(\textit{n,d}) reactions. All cross section values are expressed in milibarns. Final cross section values~$\bar{\sigma}_\car$ are assigned the statistical, systematic and total uncertainties $\Delta \sigma_\mathrm{stat}$, $\Delta \sigma_\mathrm{sys}$ and \mbox{$\Delta \sigma_\mathrm{tot}=\sqrt{(\Delta \sigma_\mathrm{stat})^2+(\Delta \sigma_\mathrm{sys})^2}$}, respectively. Overall systematic uncertainties are calculated from the model-related and flux uncertainties $\Delta \sigma_\mathrm{model}$ and $\Delta \sigma_\mathrm{flux}$ (fixed at 3\%) as \mbox{$\Delta \sigma_\mathrm{sys}=\sqrt{(\Delta \sigma_\mathrm{model})^2+(\Delta \sigma_\mathrm{flux})^2}$}. Model-related uncertainties are calculated by Eq.~(\ref{sys_uncer}). The first and the last point for each reaction (with $\bar{\sigma}_\car$ in parentheses) are considered unreliable.}
\centering

\begin{tabular}{ccccccc|cccc|ccc}
\hline\hline
\multicolumn{14}{c}{\vrt\textbf{$\boldsymbol{^{\car}}$C(\textit{n,p})}\vrt} \\
\hline
$\boldsymbol{\en}$ && $\boldsymbol{\bar{\sigma}_\car}$ & $\boldsymbol{\Delta \sigma_\mathrm{stat}}$ & $\boldsymbol{\Delta \sigma_\mathrm{sys}}$ & $\boldsymbol{\Delta \sigma_\mathrm{tot}}$ &&& $\boldsymbol{\Delta \sigma_\mathrm{talys}}$ & 
$\boldsymbol{\delta\sigma_\mathrm{max}}$ &&& $\boldsymbol{\Delta \sigma_\mathrm{model}}$ & $\boldsymbol{\Delta \sigma_\mathrm{flux}}$ \\
\hline
15 MeV && \textbf{(32.3)} & \textbf{6.4} (19.8\%) & \textbf{5.9} (18.3\%) & \textbf{8.6} (26.6\%) &&& 0.10 & 5.8 &&& 5.8 (18.0\%) & 1.0 (3.0\%) \\
16 MeV && \textbf{18.4} & \textbf{1.8} (9.8\%) & \textbf{1.2} (6.5\%) & \textbf{2.2} (12.0\%) &&& 0.01 & 1.0 &&& 1.0 (5.4\%) & 0.6 (3.0\%) \\
17 MeV && \textbf{28.4} & \textbf{1.5} (5.3\%) & \textbf{1.0} (3.5\%) & \textbf{1.8} (6.3\%) &&& 0.17 & 0.5 &&& 0.5 (1.8\%) & 0.9 (3.0\%) \\
18 MeV && \textbf{46.6} & \textbf{1.6} (3.4\%) & \textbf{3.1} (6.7\%) & \textbf{3.4} (7.3\%) &&& 0.17 & 2.7 &&& 2.7 (5.8\%)  & 1.4 (3.0\%) \\
19 MeV && \textbf{40.0} & \textbf{1.3} (3.3\%) & \textbf{4.4} (11.0\%) & \textbf{4.6} (11.5\%) &&& 0.01 & 4.2 &&& 4.2 (10.5\%) & 1.2 (3.0\%) \\
20 MeV && \textbf{53.3} & \textbf{1.5} (2.8\%) & \textbf{3.6} (6.8\%) & \textbf{3.9} (7.3\%) &&& 0.04 & 3.2 &&& 3.2 (6.0\%) & 1.6 (3.0\%) \\
21 MeV && \textbf{69.9} & \textbf{1.6} (2.3\%) & \textbf{2.8} (4.0\%) & \textbf{3.2} (4.6\%) &&& 0.09 & 1.8 &&& 1.8 (2.6\%) & 2.1 (3.0\%) \\
22 MeV && \textbf{75.3} & \textbf{1.6} (2.1\%) & \textbf{2.3} (3.1\%) & \textbf{2.8} (3.7\%) &&& 0.15 & 0.3 &&& 0.3 (0.4\%) & 2.3 (3.0\%) \\
23 MeV && \textbf{69.1} & \textbf{1.5} (2.2\%) & \textbf{2.1} (3.0\%) & \textbf{2.6} (3.8\%) &&& 0.20 & 0.3 &&& 0.3 (0.4\%) & 2.1 (3.0\%) \\
24 MeV && \textbf{67.3} & \textbf{1.4} (2.1\%) & \textbf{3.2} (4.8\%) & \textbf{3.5} (5.2\%) &&& 0.25 & 2.5 &&& 2.5 (3.7\%) & 2.0 (3.0\%) \\
25 MeV && \textbf{52.7} & \textbf{1.3} (2.5\%) & \textbf{3.3} (6.3\%) & \textbf{3.5} (6.6\%) &&& 0.16 & 2.8 &&& 2.8 (5.3\%) & 1.6 (3.0\%) \\
26 MeV && \textbf{(52.7)} & \textbf{1.1} (2.1\%) & \textbf{3.1} (5.9\%) & \textbf{3.3} (6.3\%) &&& 0.05 & 2.7 &&& 2.7 (5.1\%) & 1.6 (3.0\%) \\
\hline\hline\\
\hline\hline
\multicolumn{14}{c}{\vrt\textbf{$\boldsymbol{^{\car}}$C(\textit{n,d})}\vrt} \\
\hline
$\boldsymbol{\en}$ && $\boldsymbol{\bar{\sigma}_\car}$ & $\boldsymbol{\Delta \sigma_\mathrm{stat}}$ & $\boldsymbol{\Delta \sigma_\mathrm{sys}}$ & $\boldsymbol{\Delta \sigma_\mathrm{tot}}$ &&& $\boldsymbol{\Delta \sigma_\mathrm{talys}}$ & 
$\boldsymbol{\delta\sigma_\mathrm{max}}$ &&& $\boldsymbol{\Delta \sigma_\mathrm{model}}$ & $\boldsymbol{\Delta \sigma_\mathrm{flux}}$ \\
\hline
16 MeV && \textbf{(46.7)} & \textbf{68.6} (146.9\%) & \textbf{6.6} (14.1\%) & \textbf{68.9} (147.5\%) &&& 0.52 & 6.5 &&& 6.5 (13.9\%) & 1.4 (3.0\%) \\
17 MeV && \textbf{23.9} & \textbf{2.3} (9.6\%) & \textbf{1.7} (7.1\%) & \textbf{2.8} (11.7\%) &&& 0.41 & 1.6 &&& 1.6 (6.7\%) & 0.7 (3.0\%) \\
18 MeV && \textbf{25.4} & \textbf{1.4} (5.5\%) & \textbf{4.9} (19.3\%) & \textbf{5.1} (20.1\%) &&& 0.42 & 4.8 &&& 4.8 (18.9\%) & 0.8 (3.0\%) \\
19 MeV && \textbf{32.2} & \textbf{1.3} (4.0\%) & \textbf{5.5} (17.1\%) & \textbf{5.6} (17.4\%) &&& 0.46 & 5.4 &&& 5.4 (16.8\%) & 1.0 (3.0\%) \\
20 MeV && \textbf{47.9} & \textbf{1.4} (2.9\%) & \textbf{4.4} (9.2\%) & \textbf{4.6} (9.6\%) &&& 0.47 & 4.1 &&& 4.1 (8.6\%) & 1.4 (3.0\%) \\
21 MeV && \textbf{61.7} & \textbf{1.6} (2.6\%) & \textbf{4.9} (7.9\%) & \textbf{5.2} (8.4\%) &&& 2.09 & 4.6 &&& 4.6 (7.5\%) & 1.9 (3.0\%) \\
22 MeV && \textbf{68.6} & \textbf{1.7} (2.5\%) & \textbf{6.0} (8.7\%) & \textbf{6.2} (9.0\%) &&& 4.07 & 5.7 &&& 5.7 (8.3\%)  & 2.1 (3.0\%) \\
23 MeV && \textbf{74.9} & \textbf{1.7} (2.3\%) & \textbf{6.2} (8.3\%) & \textbf{6.5} (8.7\%) &&& 5.53 & 5.8 &&& 5.8 (7.7\%) & 2.2 (3.0\%) \\
24 MeV && \textbf{76.9} & \textbf{1.7} (2.2\%) & \textbf{6.9} (9.0\%) & \textbf{7.2} (9.4\%) &&& 6.54 & 5.2 &&& 6.5 (8.5\%) & 2.3 (3.0\%) \\
25 MeV && \textbf{71.4} & \textbf{1.7} (2.4\%) & \textbf{6.4} (9.0\%) & \textbf{6.6} (9.2\%) &&& 5.99 & 4.7 &&& 6.0 (8.4\%) & 2.1 (3.0\%) \\
26 MeV && \textbf{(65.9)} & \textbf{1.4} (2.1\%) & \textbf{6.1} (9.3\%) & \textbf{6.2} (9.4\%) &&& 4.62 & 5.7 &&& 5.7 (8.6\%) & 2.0 (3.0\%) \\
\hline\hline
\end{tabular}

\label{tab_results}
\end{table*}

The final results of our analysis are listed in Tab.~\ref{tab_results}. As discussed above, we obtain the final cross section values~$\bar{\sigma}_\car$ by averaging the analysis results from 384 TALYS model sets. We use the unweighted average, as the uncertainties from different model sets have no bearing upon the reliability of specific sets. For the same reason, we calculate the final statistical uncertainties~$\Delta \sigma_\mathrm{stat}$ from Tab.~\ref{tab_results} as the arithmetic mean of the uncertainties from 384 model sets (their variations between specific model sets being negligible in any case).

For each reaction the first and the last point from Tab.~\ref{tab_results} (with a value of $\bar{\sigma}_\car$ in parentheses) are unreliable. We report them here only for completeness. The first point is at the limit of the lowest detection threshold, where the detection efficiency and the acquired statistics are very low (see Figs.~\ref{fig_nd_all} and~\ref{fig_counts}). The last point is outside our targeted range (see Section~\ref{part_discrimination}). It has been treated differently from other points and should not be considered on equal footing as the points below 26~MeV.

\begin{figure}[t!]
\centering
\includegraphics[width=1\linewidth]{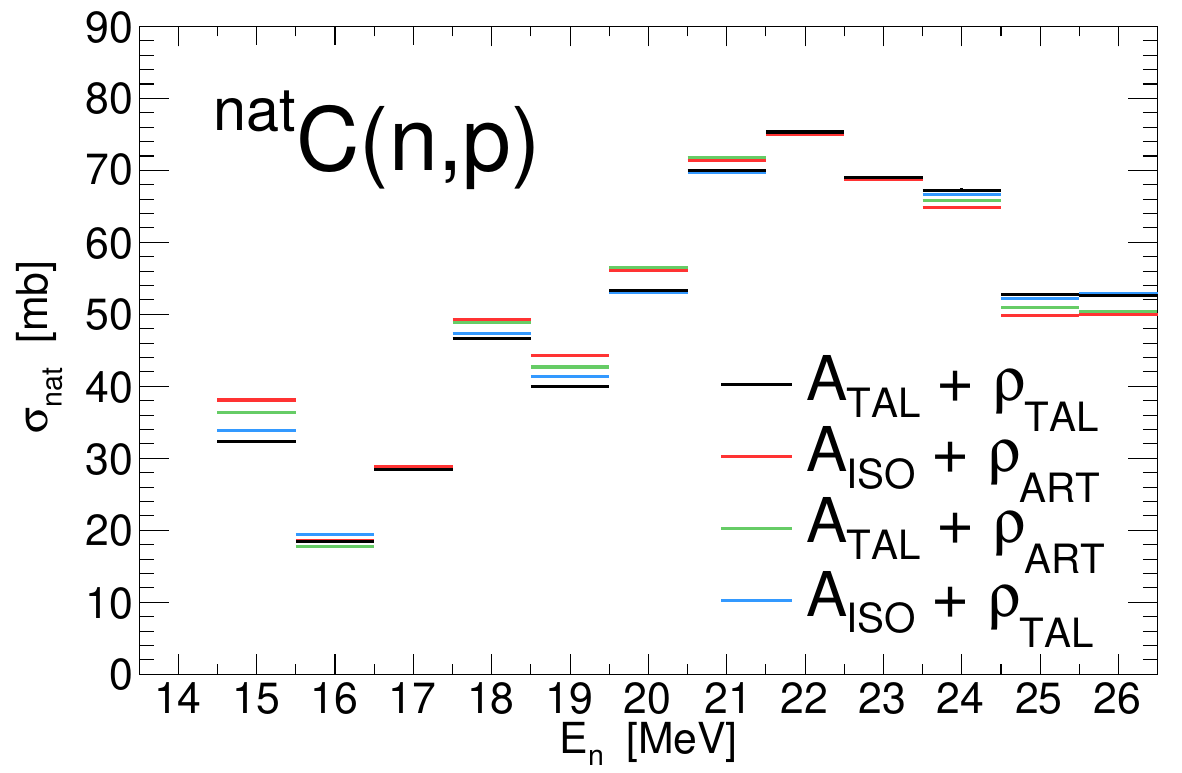}
\includegraphics[width=1\linewidth]{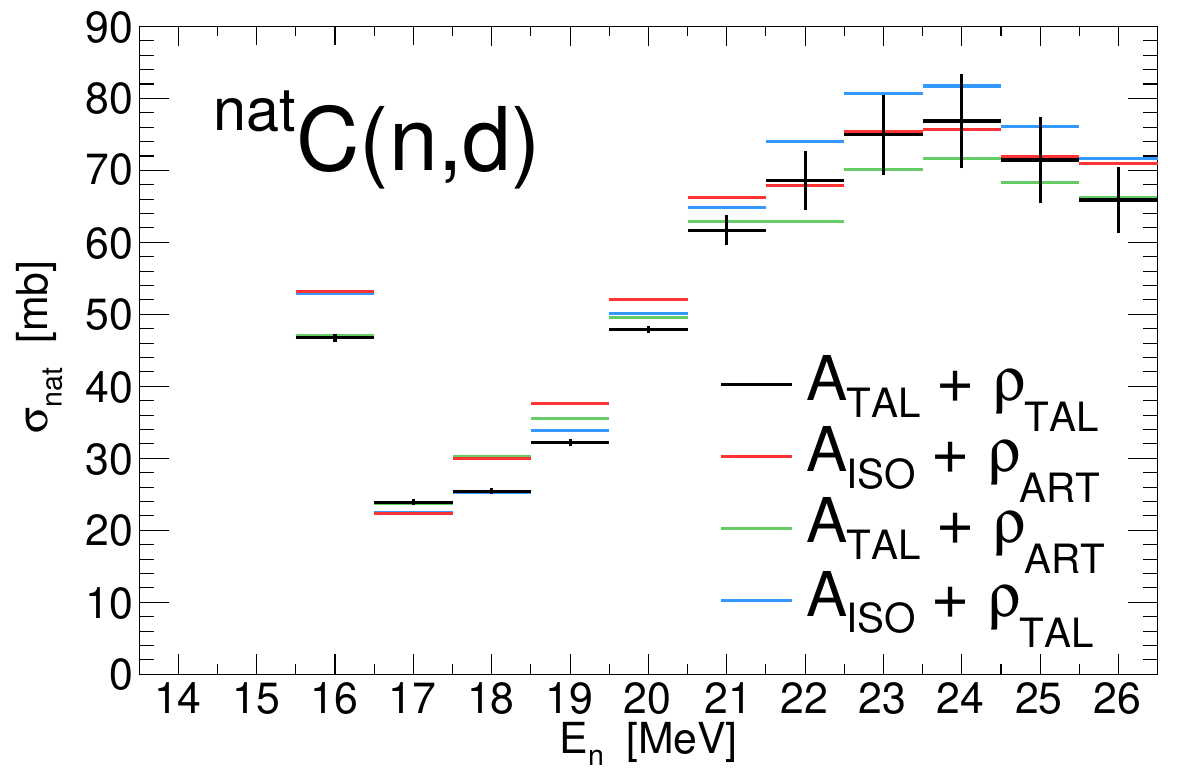}
\vspace*{-4mm}
\caption{Analysis results obtained from different combinations of angular distributions ($A_\mathrm{TAL}$ from TALYS or isotropic $A_\mathrm{ISO}$) and branching ratios ($\rho_\mathrm{TAL}$ from TALYS or artificial $\rho_\mathrm{ART}$). Error-bars are shown only for TALYS results and correspond to their RMS ($\Delta\sigma_\mathrm{talys}$ from Tab.~\ref{tab_results}).}
\label{fig_vars}
\vspace*{-2mm}
\end{figure}

In estimating the model-related uncertainties~$\Delta \sigma_\mathrm{model}$ we use the following conservative procedure. For each energy bin we first observe the RMS of the results obtained with 384 model sets. These values are listed in Tab.~\ref{tab_results} under~$\Delta \sigma_\mathrm{talys}$. Alongside 384 repeated analyses using separate model sets, we also perform three additional analyses. In one analysis variant we use fully artificial distributions from Section~\ref{systematic}: isotropic angular distributions and artificial branching ratios from Eq.~(\ref{br_art}). Furthermore, we combine the TALYS data with the artificial distributions, performing 384 additional analyses with isotropic angular distributions and TALYS branching ratios, together with 384 analyses using TALYS angular distributions and artificial branching ratios. We treat these two additional sets of 384 analyses in the same way as a set of 384 TALYS analyses and show the obtained results in Fig.~\ref{fig_vars}. For three considered analysis variants using artificial distributions (denoted by~\mbox{$i=1,2,3$}) we determine a maximum deviation~$\delta\sigma_\mathrm{max}$ from all-TALYS (\mbox{$A_\mathrm{TAL}+\rho_\mathrm{TAL}$} in Fig.~\ref{fig_vars}) values~$\bar{\sigma}_\car$:
\begin{linenomath}\begin{equation}
\delta\sigma_\mathrm{max}=\max_{i} \big|\bar{\sigma}_\car-\bar{\sigma}_i\big|,
\end{equation}\end{linenomath}
and conservatively estimate the model-related uncertainty in our data as a maximum between the RMS of TALYS-variations and~$\delta\sigma_\mathrm{max}$:
\begin{linenomath}\begin{equation}
\Delta \sigma_\mathrm{model}=\max \big(\Delta \sigma_\mathrm{talys} \,,\, \delta\sigma_\mathrm{max} \big).
\label{sys_uncer}
\end{equation}\end{linenomath}
For clarity, Fig.~\ref{fig_vars} shows the error-bars only for all-TALYS results~$\bar{\sigma}_\car$, which correspond to the RMS of variations between TALYS model sets ($\Delta \sigma_\mathrm{talys}$). Most of the error-bars are so small, especially for the $^\car$C(\textit{n,p}) reaction, that they are barely visible.

\begin{figure}[t!]
\centering
\includegraphics[width=1\linewidth]{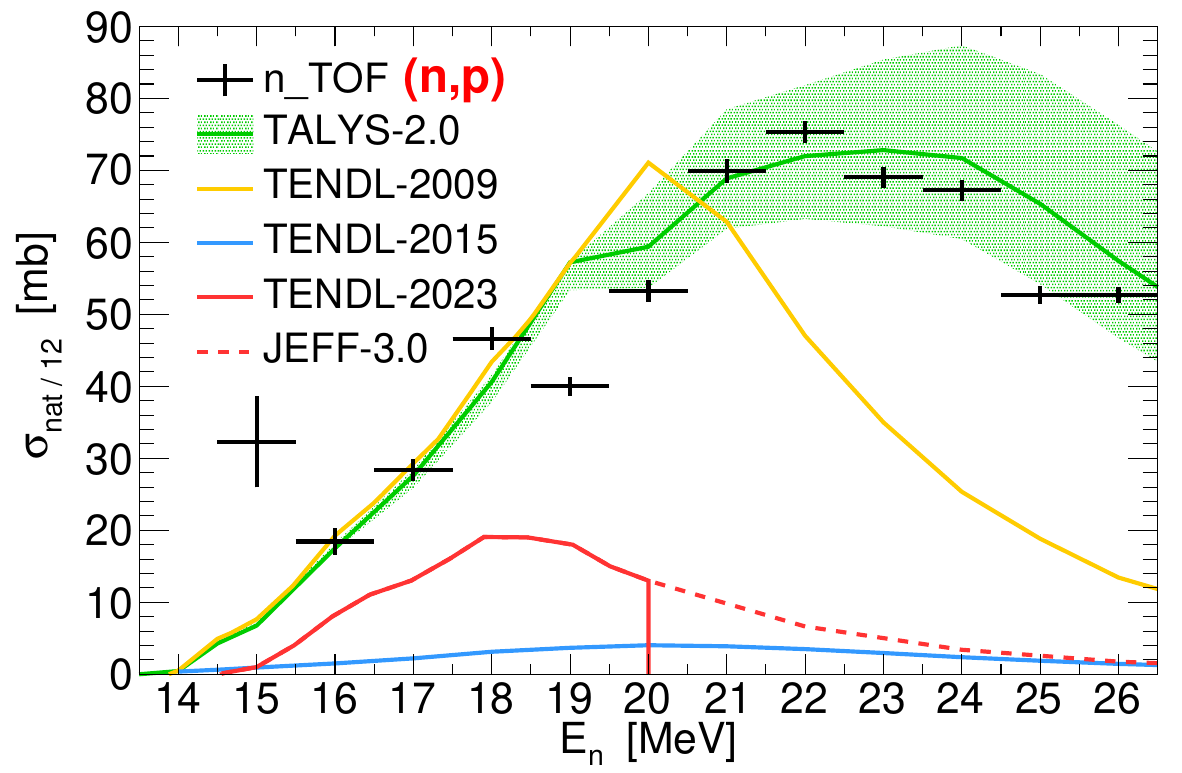}
\includegraphics[width=1\linewidth]{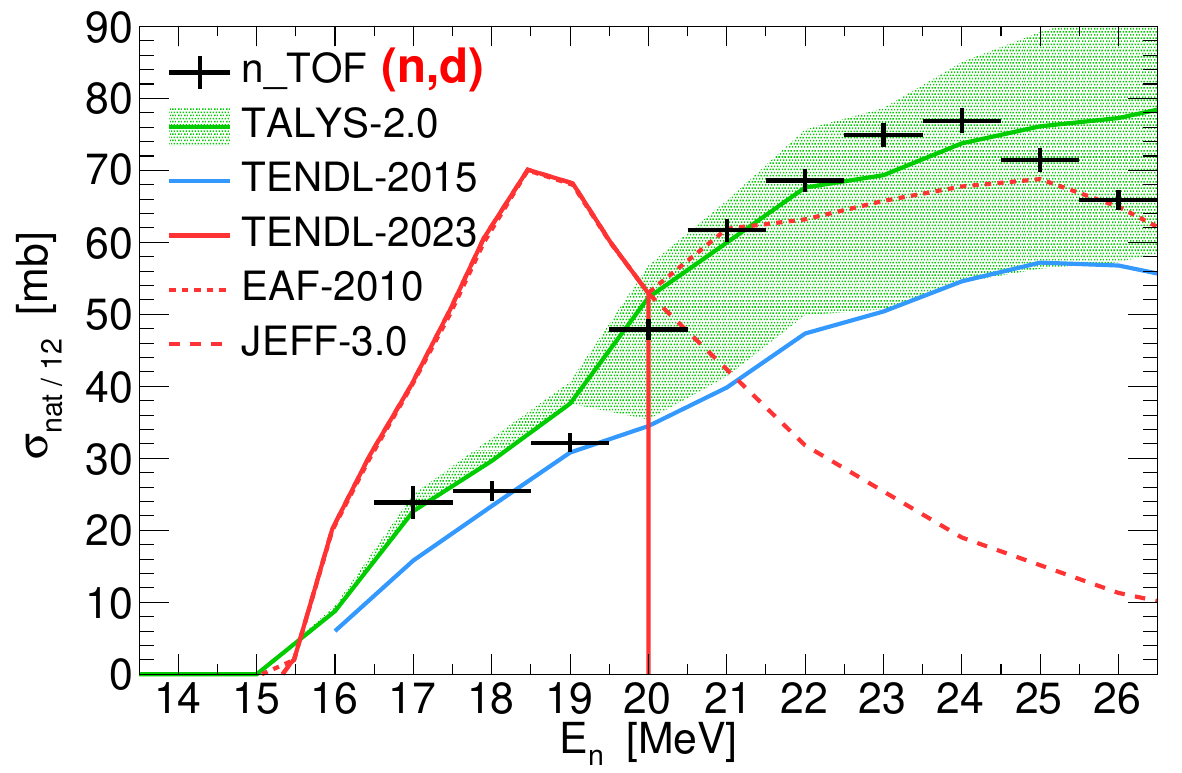}
\vspace*{-5mm}
\caption{The $^\car$C(\textit{n,p}) and $^\car$C(\textit{n,d}) cross sections from n\_TOF, compared with relevant releases of the TENDL evaluation library. \mbox{JEFF-3.0} and \mbox{EAF-2010} libraries provide extensions of \mbox{TENDL-2023} evaluation, which is representative of most of the latest evaluation libraries (some libraries list their evaluations under $^\car$C, others under $^{12}$C). Shaded area shows a range of \mbox{TALYS-2.0} cross sections from 480 investigated model sets. Full TALYS lines show the model predictions in best agreement with the n\_TOF data (see the main text for details). Error-bars show only the statistical uncertainties in the n\_TOF data ($\Delta\sigma_\mathrm{stat}$ from Tab.~\ref{tab_results}). Due to the large statistical uncertainty, a point at 16~MeV for the (\textit{n,d}) reaction has been excluded for display purposes.}
\label{fig_tendl}
\vspace*{-3mm}
\end{figure}

The evaluated neutron flux $\phi(\en)$ from Eq.~(\ref{kernel}) is the second major source of systematic uncertainties, alongside the model-related ones. Above 10~MeV the n\_TOF flux is measured using Parallel Plate Avalanche Counters relying on the $^{235}$U(\textit{n,f}) reaction~\cite{flux_ear1}. The systematic uncertainty in the flux normalization in this energy range amounts to 3\%, directly propagating to the uncertainties in the extracted cross sections. We therefore assign to each data point from Tab.~\ref{tab_results} a flux-related uncertainty \mbox{$\Delta \sigma_\mathrm{flux}=0.03\bar{\sigma}_\car$}, contributing to the overall systematic uncertainty as \mbox{$(\Delta \sigma_\mathrm{sys})^2=(\Delta \sigma_\mathrm{model})^2+(\Delta \sigma_\mathrm{flux})^2$}.

Figure~\ref{fig_tendl} compares the final n\_TOF results with several data evaluations or calculations, showing the error bars that correspond to the reported statistical uncertainties~$\Delta \sigma_\mathrm{stat}$. \mbox{TENDL-2023} evaluation is representative of most other modern evaluation libraries, such as ENDF/B-VIII.1. Some libraries list their evaluations under $^\car$C, others under $^{12}$C. Displayed \mbox{JEFF-3.0} (same as BROND-2.2) and \mbox{EAF-2010} evaluations provide extensions of \mbox{TENDL-2023} evaluations above 20~MeV (later versions JEFF-3.3 and BROND-3.1 stop at 20~MeV). 

For $^\car$C(\textit{n,p}) there is a large discrepancy between the n\_TOF data and most major libraries, represented here by the TENDL-2023 evaluation. A notable exception is the TENDL-2009 library below 20~MeV, based on TALYS-1.2 calculations~\cite{tendl_2009}. On the other hand, while TENDL-2015 provides by far the worst agreement with the (\textit{n,p}) data from n\_TOF, the opposite is true for the (\textit{n,d}) data, which seem to be well modelled by this library. EAF-2010 evaluation is in a good agreement with the (\textit{n,d}) data above 20~MeV.

Shaded areas from Fig.~\ref{fig_tendl} show a range of \mbox{TALYS-2.0} cross sections from 480 models sets described in Section~\ref{talys_input}. Full lines going through these shaded areas show the specific TALYS cross sections in best agreement with the n\_TOF data. This optimal agreement is obtained with the following parameters values from Tab.~\ref{tab_models}: \mbox{ldmodel=2}, \mbox{jlmomp=y} and \mbox{colenhance=y} for the (\textit{n,p}) reaction; \mbox{ldmodel=1}, \mbox{jlmomp=n} and \mbox{colenhance=n} for the (\textit{n,d}) reaction. The rest of the parameters negligibly affect these optimal cross section dependences. Hence, the full TALYS lines show the cross sections from some arbitrarily selected model sets determined by these three relevant parameters. It should be stressed that the absolute values of TALYS cross sections were at no point used in the analysis of the experimental data; only the \textit{relative} values (i.e. the branching ratios) were used for separate excited states in the relevant boron nuclei. Therefore, a remarkable agreement between the n\_TOF results and \mbox{TALYS-2.0} calculations provides a strong indication on the reliability of TALYS calculations.

\begin{figure}[t!]
\centering
\includegraphics[width=1\linewidth]{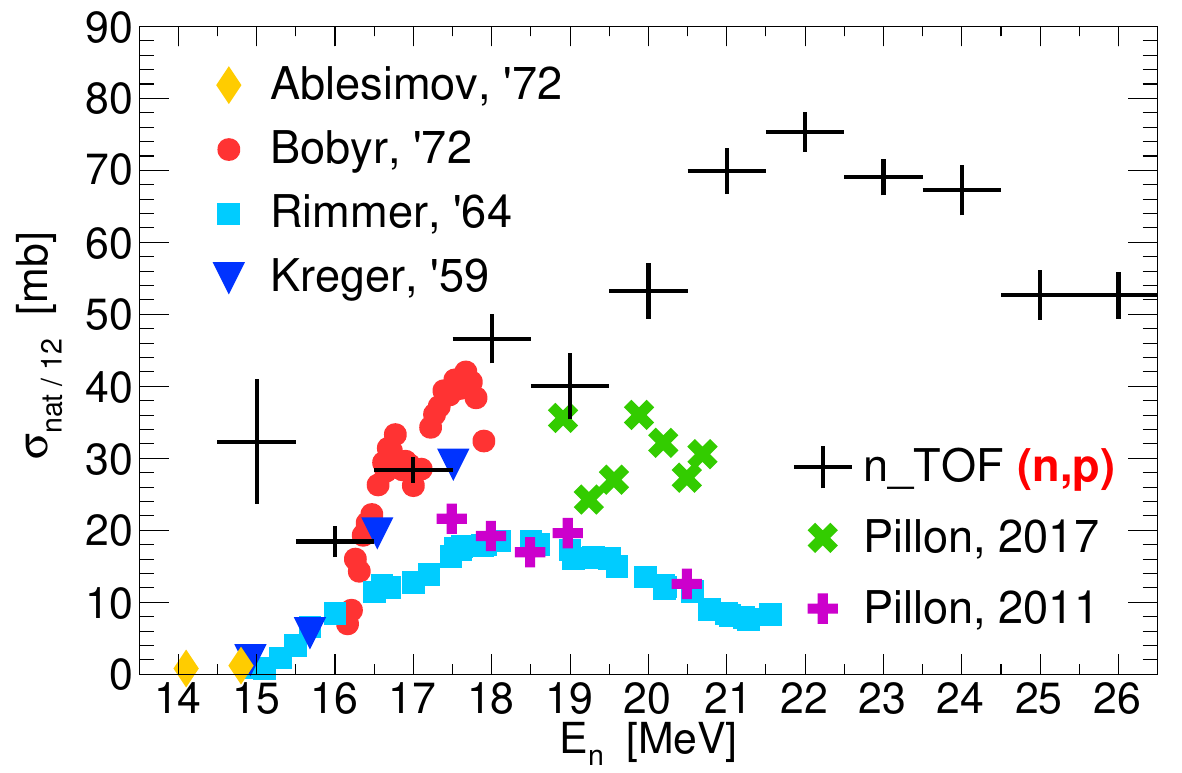}
\includegraphics[width=1\linewidth]{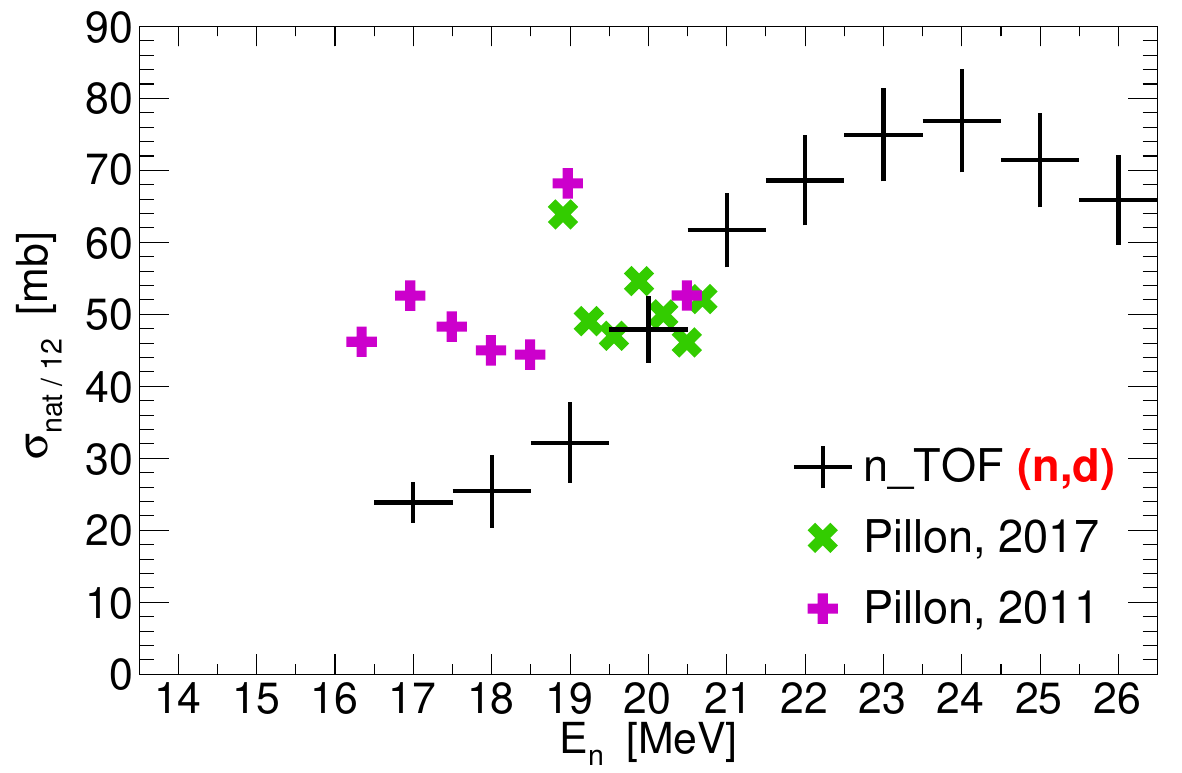}
\vspace*{-4.5mm}
\caption{n\_TOF cross sections compared with past experimental data from \cite{kreger_1959,rimmer_1968,ablesimov_1971,bobyr_1972,pillon_2011,pillon_2017}. As in Fig.~\ref{fig_tendl}, the 16~MeV (\textit{n,d}) point is not displayed. Total uncertainties $\Delta \sigma_\mathrm{tot}$ are shown.} 
\label{fig_exfor}
\vspace*{-3mm}
\end{figure}

Figure~\ref{fig_exfor} compares the n\_TOF data with the past experimental results from Kreger and Kern~\cite{kreger_1959}, Rimmer and Fisher~\cite{rimmer_1968}, Ablesimov et al.~\cite{ablesimov_1971}, Bobyr et al.~\cite{bobyr_1972} and Pillon et al.~\cite{pillon_2011,pillon_2017}. The error-bars show a total uncertainty $\Delta \sigma_\mathrm{tot}$ of the n\_TOF points.

The n\_TOF results agree well with the (\textit{n,p}) data from Kreger and Kern~\cite{kreger_1959} and Bobyr et al.~\cite{bobyr_1972}. For both the (\textit{n,p}) and (\textit{n,d}) reactions both datasets from Pillon et al.~\cite{pillon_2011,pillon_2017} provide only the partial cross sections for the lowest energy states in the daughter boron nuclei. However, within $\approx$4~MeV from the reaction threshold the cross section is fully determined by these states and can thus be compared to the n\_TOF data. There seems to be some agreement for the (\textit{n,d}) reaction around 20~MeV, but it is rather questionable due to the inconsistencies in the (\textit{n,p}) data from the two publications by Pillon et al. The (\textit{n,p}) data from Rimmer and Fisher~\cite{rimmer_1968} -- as the basis for ENDF/B-VIII.1 and TENDL-2023 (see Fig.~\ref{fig_tendl}) -- are clearly incompatible with the n\_TOF results. The recent measurements by Majerle et al.~\cite{majerle_2020}, Kuvin et al.~\cite{kuvin_2021} and Wantz et al.~\cite{wantz_2025} are not directly comparable to the present data, as they yield the partial cross sections only for the ground state or the first excited state of the residual boron nuclei.

The $^\car$C(\textit{n,p}) results reported here provide wider energy dependence, higher cross section and higher integral value up to 25~MeV than any of the previously available experimental datasets. This finding is consistent with the earlier integral cross section measurement from n\_TOF~\cite{carbon_prc,carbon_epja}, based on a completely different experimental procedure -- an activation measurement based on a detection of $\beta$ rays from a decay of $^{12}$B nuclei by means of the liquid C$_6$D$_6$ scintillators (indeed, the unexpectedly high integral value found therein motivated the present study). Considering the substantial differences in the experimental procedures between the earlier integral and the present energy-differential measurement, the latest $^\car$C(\textit{n,p}) results from n\_TOF may be regarded as an independent confirmation of this finding. As such, the results of this work call for further experimental and theoretical study of the challenging (\textit{n},cp) reactions on carbon.

\vspace*{-4mm}

\section{Conclusions}
\label{conclusions}

\vspace*{-1mm}

The energy differential cross sections for the $^\car$C(\textit{n,p}) and $^\car$C(\textit{n,d}) reactions have been measured at the first experimental area (EAR1) of the n\_TOF facility at CERN, from the particle detection threshold up to 25~MeV. The reaction thresholds of $\approx$14~MeV for the (\textit{n,p}) and $\approx$15~MeV for the (\textit{n,d}) reaction are determined by the lower threshold between the two carbon isotopes ($^{12}$C and $^{13}$C) from its natural abundance. The reported cross section data start approximately 2~MeV above the reaction thresholds -- at 16~MeV for (\textit{n,p}) and 17~MeV for (\textit{n,d}). Though the measured data extend up to 180~MeV, the upper limit of the reported cross sections is determined by the ability to reliably discriminate between the protons and deuterons from the~(\textit{n,p}) and~(\textit{n,d}) reactions, and between the charged particles from the competing channels, such as protons from the (\textit{n,np}) reaction.

Two separate silicon telescopes were used, placed in suitably chosen geometric configuration so as to cover a large angular range. The telescopes consist of separate silicon strips allowing for the $\Delta E$-$E$ technique to be used for particle discrimination. The energy of the incident neutrons was determined by a time-of-flight technique. The charged particle discrimination was performed by taking advantage of the neural networks carefully optimized for each relevant $\Delta E$-$E$ pair of silicon strips. The total of 62 pairs of strips were used in the final data analysis. Though the placement of silicon strips around the sample allows for some sensitivity to the angular distribution of the reaction products, the obtained statistics was insufficient for its reliable extraction.

The energy and angular distributions of emitted protons and deuterons are sensitive to a rate of population of the energetically available excited states of the daughter nuclei, i.e. the boron isotopes $^{11}$B, $^{12}$B, $^{13}$B. While the detection efficiencies were determined by the dedicated Geant4 simulations of the experimental setup, an external source of information was required for the estimation of the angular distributions and the energy-dependent branching ratios for the excited states in boron nuclei. TALYS-2.0 calculations were used as a main source of these distributions. An in-depth investigation of the systematic model uncertainties was performed, based on the repeated analyses employing the data from 384 different model sets from TALYS. The artificial set of branching ratios and angular distributions was also considered. The systematic effects of adopting these model predictions was found to be below 10\% for the $^\car$C(\textit{n,p}) reaction, and below 20\% for the $^\car$C(\textit{n,d}) reaction. A special care was also taken to account for the finer experimental effects, such as the resolution function of the neutron beam. A systematic uncertainty in the normalization of the neutron flux was considered as well.

Final cross section results were extracted as averages over the natural abundances of $^{12}$C and $^{13}$C, yielding the data for the natural carbon, already published as a Letter in Ref.~\cite{carbon_letter}. These results may also be considered as representative of the (\textit{n,p}) and (\textit{n,d}) reactions on $^{12}$C, due to its high natural abundance of 98.9\% and the estimated similarity of its cross sections to those for $^{13}$C. The results were compared to the available experimental data and evaluation libraries (the majority of present libraries are represented in this work by TENDL-2023). A significant disagreement was found with most of the available libraries, for both the (\textit{n,p}) and (\textit{n,d}) reaction. On the other hand, a good agreement was obtained with the TALYS-2.0 calculations. It should be stressed that the absolute values of the TALYS cross sections were at no point used in the analysis of the experimental data. The agreement with the past experimental data is mixed, as they are discrepant among themselves and generally cover a narrower energy range  than the present n\_TOF data. However, our experimental findings for the (\textit{n,p}) reaction  are consistent with the old measurements by Kreger and Kern~\cite{kreger_1959}, and by Bobyr et al.~\cite{bobyr_1972}. Finally, the present energy-differential results seem to be consistent with the earlier  $^{12}$C(\textit{n,p}) integral measurement  from n\_TOF, based on an entirely different experimental technique~\cite{carbon_prc,carbon_epja}. Both n\_TOF measurements clearly point to a cross section higher than reported in the evaluation libraries and the majority of previous experimental data. Due to the obviously challenging nature of the measurement and the importance of the (\textit{n},cp) reactions on carbon -- in nuclear medicine, diamond detectors design, fundamental nuclear physics, etc. -- the latest results from n\_TOF strongly motivate further experimental and theoretical investigation of these reactions.


\section*{ACKNOWLEDGMENTS}


This work was supported by the Croatian Science Foundation under the project number HRZZ-IP-2022-10-3878, and by the National Science Centre in Poland (Grant No. UMO-2021/41/B/ST2/00326). This project has received funding from the European Union’s Horizon Europe Research and Innovation programme under Grant Agreement No 101057511. D.~Rochman acknowledges support from EU APRENDE project, number 101164596. Geant4 simulations were run at the Laboratory for Advanced Computing, Faculty of Science, University of Zagreb.

\appendix*

\section{Overall detector efficiency}

We discuss here the \textit{overall} detection efficiency of the entire detection setup consisting of two silicon telescopes. By the detection efficiencies we consider the true probabilities for particles to be detected, once they have been emitted. The efficiencies for particular $\Delta E$-$E$ pairs of silicon strips are accurately determined by the \mbox{$\varepsilon_{\Pij,\C}(\x,\en,\cos\theta)$} terms from Eq.~(\ref{kernel}), covered in more detail in Ref.~\cite{angular}. These are, therefore, the probabilities for the reaction products (protons or deuterons) to be detected once they have been emitted with a specific energy (determined by the incident neutron energy~$\en$, the residual nucleus excited state~$\x$ and the emission angle~$\theta$), averaged over the depth and breadth of the carbon sample. They not only account for the angular coverage of each particular pair of silicon strips (e.g., see Fig.~2 from Ref.~\cite{angular}), but also for the particle loss due to their loss of energy on their way through the carbon sample and the silicon $\Delta E$-layer towards the final $E$-layer.

In order to obtain the neutron energy dependence of the detection efficiency of the entire experimental setup, we need to integrate the angle-differential efficiencies over the entire solid angle around the carbon sample, taking into account the angular distribution of the reaction products. We also need to take into account the two carbon isotopes as the separate sources of the reaction products, and a distribution of products over the available excited states in daughter nuclei. Using the quantities defined in Section~\ref{modeling} -- crucial to the modeling of Eqs.~(\ref{master}) and~(\ref{kernel}) -- the overall efficiency $\bar{\varepsilon}(\en)$ for \textit{any} reaction product of a given type (produced by the neutrons of energy~$\en$) takes the form:
\begin{widetext}
\begin{linenomath}\begin{equation}
\bar{\varepsilon}(\en)=\frac{\sum_\C \alpha_\C\sigma_\C(\en) \sum_\x\rho_\C(\x,\en) \int_{-1}^1\D(\cos\theta)\times A_\C(\x,\en,\cos\theta) \sum_\Pij \varepsilon_{\Pij,\C}(\x,\en,\cos\theta) }{\sum_\C \alpha_\C\sigma_\C(\en)},
\label{tot_eps}
\end{equation}\end{linenomath}
\end{widetext}
where we have introduced the carbon abundances \mbox{$\alpha_{12}=98.9\%$} and \mbox{$\alpha_{13}=1.1\%$},  directly related to the previously used areal densities~$\eta_\C$ as \mbox{$\alpha_\C=\eta_\C/(\eta_{12}+\eta_{13})$}. One such expression, therefore, holds for the protons and the separate one for the deuterons. Clearly, one would need to know in advance all the angular distributions \mbox{$A_\C(\x,\en,\cos\theta)$}, branching ratios \mbox{$\rho_\C(\x,\en)$} and the cross sections \mbox{$\sigma_\C(\en)$} in order to accurately determine $\bar{\varepsilon}(\en)$. Thus -- unlike the differential efficiencies \mbox{$\varepsilon_{\Pij,\C}(\x,\en,\cos\theta)$} -- the overall efficiency is no longer an intrinsic property the experimental setup; it also reflects the nature of a specific reaction being measured.

One of the central points of the data analysis from this work has been that the required terms \mbox{$A_\C,\rho_\C,\sigma_\C$} are not well established. However, based on some simplified assumptions one can get a single reasonable estimate of $\bar{\varepsilon}(\en)$ dependence (instead of multiple ones, for each particular model set of TALYS parameters from Table~\ref{tab_models}). This estimate is, of course, not to be used for an accurate data analysis, but only as a rough demonstration of a response of the experimental setup. In that, all angular distributions are best assumed isotropic. The cross sections $\sigma_{12}$ and $\sigma_{13}$ can be approximated as being equal, the effect of this procedure being justified by the dominance of one isotope's abundance over another (\mbox{$\alpha_{12}\gg\alpha_{13}$}). They are sufficiently similar for  this purpose, as demonstrated by Fig.~\ref{fig_13c}. For the branching ratios there are no such straightforward options at hand. To this end we use the artificially constructed set from Eq.~(\ref{br_art}), so as to avoid committing to any particular TALYS set. Recognizing the portion of Eq.~(\ref{tot_eps}) starting from the integral sign as the overall efficiency \mbox{$\tilde{\varepsilon}_\C(\x,\en)$} for the detection of reaction products leaving the residual nucleus in the $\x$-th excited state, after these assumptions one is left with:
\begin{linenomath}\begin{equation}
\bar{\varepsilon}(\en)=\sum_\C \alpha_\C \sum_\x\rho_\C(\x,\en) \tilde{\varepsilon}_\C(\x,\en).
\label{eps_approx}
\end{equation}\end{linenomath}

\begin{figure}[b!]
\centering
\includegraphics[width=1\linewidth]{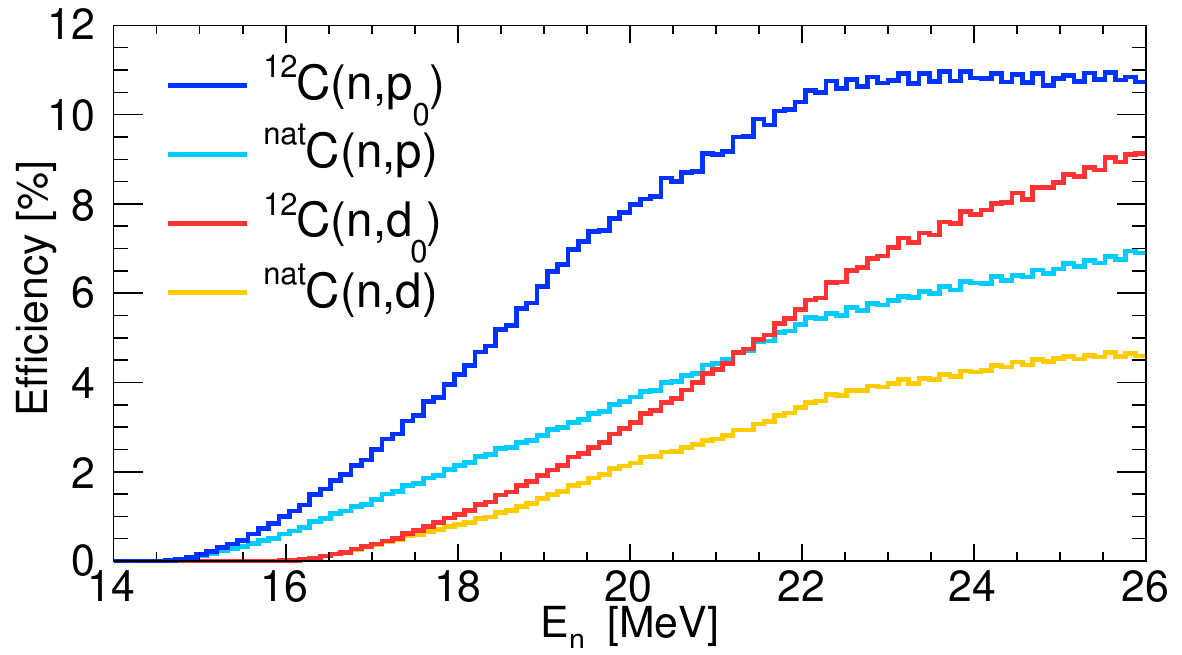}
\caption{Detection efficiencies \mbox{$\tilde{\varepsilon}_{12}(0,\en)$} for the protons and deuterons from $^{12}$C leaving the residual boron nuclei in the ground state, together with the overall efficiencies $\bar{\varepsilon}(\en)$ from Eq.~(\ref{eps_approx}), averaged over the excited states in boron nuclei and over the natural abundances of $^{12}$C, $^{13}$C isotopes.}
\label{fig_efficiency}
\end{figure}

As a starting point, Fig.~\ref{fig_efficiency} shows the efficiencies \mbox{$\tilde{\varepsilon}_{12}(0,\en)$} related to the \mbox{$^{12}$C(\textit{n,p$_0$})} and \mbox{$^{12}$C(\textit{n,d$_0$})} reactions -- i.e. for the protons and deuterons from the dominant $^{12}$C leaving the residual boron nuclei in the ground state -- because these reaction channels open first and produce the highest-energetic particles. Fig.~\ref{fig_efficiency} also shows the overall $\bar{\varepsilon}(\en)$ from Eq.~(\ref{eps_approx}) for the $^\car$C(\textit{n,p}) and $^\car$C(\textit{n,d}) reactions, accounting for both carbon isotopes and all the excited states (the results are basically the same as for the $^{12}$C alone: \mbox{$\bar{\varepsilon}(\en)\approx \sum_\x\rho_{12}(\x,\en) \tilde{\varepsilon}_{12}(\x,\en)$}, due to \mbox{$\alpha_{12}\gg\alpha_{13}$}).

A plot for the \mbox{$^{12}$C(\textit{n,p$_0$})} reaction shows that from 22~MeV the efficiency for the \textit{p$_0$} protons saturates around 11\%. At lower neutron energies the \textit{p$_0$} protons that manage to reach the $E$-layer (losing a certain amount of energy in the $\Delta E$-layer) have a shorter range in the carbon sample than its thickness, so the certain amount of protons from the ``far end'' of the sample is lost to the detection in the $E$-layer. With increasing neutron energy the protons from the greater sample depths manage to reach the $E$-layer, thus increasing a relative portion of the detected ones. This continues until the protons from the ``farthest end'' can be detected, thus saturating the efficiency. For deuterons from the \mbox{$^{12}$C(\textit{n,d$_0$})} reaction this saturation does not yet occur below 26~MeV.

A reduction in the efficiency for detecting \textit{any} proton or deuteron from the $^\car$C(\textit{n,p}) and $^\car$C(\textit{n,d}) reactions is due to the activation of the higher excited states in residual boron nuclei at higher neutron energies. Compared to the ground state, these states lead to the emission of less energetic products which are then more efficiently stopped before reaching the $E$-layer. For this reason the overall efficiency is almost halved and its saturation is not even fully achieved below 26~MeV.

This efficiency behaviour is the reason why the sample thickness of 0.25~mm was selected. Thinner samples decrease a detection yield around 26~MeV (the upper edge of a targeted energy range), where protons from the first few reaction channels do saturate. Thicker samples, however, do not lead to an appreciable increase in the overall efficiency, which almost saturates around 26~MeV. Hence, 0.25~mm was found to be an optimum between the maximization of a detection yield, and the minimization of the multiple scattering effects and of the possible background from the competing reactions.


\end{document}